\documentclass[aps,prx,superscriptaddress,amsmath,amssymb,twocolumn,showpacs,floatfix,reprint]{revtex4-2}
\usepackage[colorlinks=true, urlcolor=blue, linkcolor=blue, citecolor=blue, pdftex]{hyperref}
\usepackage[capitalise]{cleveref}
\usepackage{xr}
\usepackage{multirow}
\usepackage[dvipsnames]{xcolor}
\usepackage[utf8]{inputenc}
\usepackage{braket}
\usepackage{graphicx}

\definecolor{C0}{HTML}{1f77b4}
\definecolor{C1}{HTML}{ff7f0e}
\definecolor{C2}{HTML}{2ca02c}
\definecolor{C3}{HTML}{d62728}
\definecolor{C4}{HTML}{9467bd}
\definecolor{C5}{HTML}{8c564b}

\let\b\boldsymbol

\begin{document}

\title{Approaching the Thermodynamic Limit with Neural-Network Quantum States}
\date{\today}

\begin{abstract}
Accessing the thermodynamic-limit properties of strongly correlated quantum matter requires simulations on very large lattices, a regime that remains challenging for numerical methods, especially in frustrated two-dimensional systems. We introduce the \textit{Spatial Attention mechanism}, a minimal and physically interpretable inductive bias for Neural-Network Quantum States, implemented as a single learned length scale within the Transformer architecture. This bias stabilizes large-scale optimization and enables access to thermodynamic-limit physics through highly accurate simulations on unprecedented system sizes within the Variational Monte Carlo framework. Applied to the spin-$\tfrac12$ triangular-lattice Heisenberg antiferromagnet, our approach achieves state-of-the-art results on clusters of up to $42\times42$ sites. The ability to simulate such large systems allows controlled finite-size scaling of energies and order parameters, enabling the extraction of experimentally relevant quantities such as spin-wave velocities and uniform susceptibilities. In turn, we find extrapolated thermodynamic limit energies systematically better than those obtained with tensor-network approaches such as iPEPS. The resulting magnetization is strongly renormalized, $\mathcal{M}_0=0.148(1)$ (about $30\%$ of the classical value), revealing that less accurate variational states systematically overestimate magnetic order. Analysis of the optimized wave function further suggests an intrinsically non-local sign structure, indicating that the sign problem cannot be removed by local basis transformations. We finally demonstrate the generality of the method by obtaining state-of-the-art energies for a $J_1$-$J_2$ Heisenberg model on a $20\times20$ square lattice, outperforming Residual Convolutional Neural Networks.
\end{abstract}

\author{Luciano Loris Viteritti}
\thanks{These authors contributed equally. Correspondence should be addressed to rrende@flatironinstitute.org and luciano.viteritti@epfl.ch}
\affiliation{Institute of Physics, \'{E}cole Polytechnique F\'{e}d\'{e}rale de Lausanne (EPFL), CH-1015 Lausanne, Switzerland}

\author{Riccardo Rende}
\thanks{These authors contributed equally. Correspondence should be addressed to rrende@flatironinstitute.org and luciano.viteritti@epfl.ch}
\affiliation{Center for Computational Quantum Physics, Flatiron Institute, 162 5th Avenue, New York, NY 10010}

\author{Subir Sachdev}
\affiliation{Center for Computational Quantum Physics, Flatiron Institute, 162 5th Avenue, New York, NY 10010}
\affiliation{Department of Physics, Harvard University, Cambridge MA 02138, USA}

\author{Giuseppe Carleo}
\affiliation{Institute of Physics, \'{E}cole Polytechnique F\'{e}d\'{e}rale de Lausanne (EPFL), CH-1015 Lausanne, Switzerland}

\maketitle
\section{Introduction}
Extracting thermodynamic-limit properties of strongly correlated quantum systems is a central goal in condensed matter physics~\cite{Anderson1987,SachdevBook}: reaching sufficiently large lattices is essential for performing controlled finite-size scaling and reliably extracting thermodynamic-limit quantities such as order parameters, excitation gaps, or correlation lengths~\cite{NeubergerZiman1989,Sandvik2010}. Despite major progress over the past decades, existing numerical methods struggle to balance accuracy and scalability, especially for frustrated systems in two dimensions, where sign problems or entanglement constraints make large-scale calculations prohibitive~\cite{TroyerWiese2005,Balents2010,EisertCramerPlenio2010_Entanglement,Schollwock2011}. 

Among the many approaches developed to tackle this problem, Neural-Network Quantum States (NQS)~\cite{carleo2017, lange2024} have emerged as a particularly promising class of variational Ans\"atze~\cite{chen2024empowering, viteritti2023prl, viteritti2025prb, roth2023}. By leveraging the expressive power of modern deep-learning architectures, NQS can efficiently represent highly entangled quantum states~\cite{Sharir2019, Sharir2022, gao2017} and capture complicated sign structures that are inaccessible to conventional methods~\cite{szabo2020, viteritti2022, rende2024stochastic, chen2024empowering}. 

\begin{figure*}[ht]
    \begin{center}
\centerline{\includegraphics[width=2.3\columnwidth]{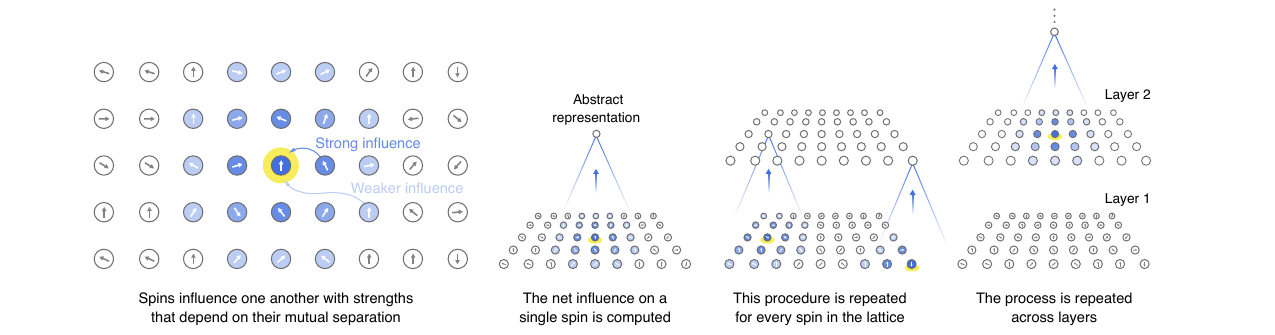}}
        \caption{\label{fig:pictorial_representation}\textbf{Schematic of the Spatial Attention mechanism}. For each input spin, an abstract representation is formed by aggregating contributions from all other spins, weighted by their spatial separation: nearby spins contribute more strongly, while contributions from distant sites are progressively suppressed. This operation is performed independently at every lattice site, yielding a new representation for each spin. The procedure is iterated across successive Transformer layers.}
    \end{center}
\end{figure*}

As in broader machine learning, the success of NQS relies on \emph{inductive biases}~\cite{inductive_bias,convit}: architectural priors that guide learning toward physically relevant solutions using finite data and noisy gradients. This naturally raises the question: \emph{which inductive biases are most relevant for variational descriptions of ground states?} Physical Hamiltonians typically involve local interactions, which constrain how correlations decay and how information propagates~\cite{LiebRobinson1972}. This behavior is formalized by the \textit{cluster property}: correlations between operators vanish as their separation, either in space or in time, increases~\cite{HastingsKoma2006}. Importantly, the locality of the Hamiltonian does \emph{not} imply short-ranged correlations in the corresponding ground states: gapless phases and critical points display algebraic correlations and long-range entanglement. This distinction is crucial for designing a proper variational state. A locality-inspired bias can therefore be valuable as an optimization prior without imposing hard representational constraints such as finite correlation length or area-law truncation.

Here, we propose an inductive bias for Transformer-based NQS inspired by the cluster property. This bias is realized through the \textit{Spatial Attention mechanism}, a minimal and physically motivated modification of the attention mechanism~\cite{viteritti2023prl,viteritti2025prb,czischek2023}, in which attention weights are reweighted by an explicit distance-dependent kernel controlled by a single length-scale parameter (see \cref{sec:att_weights}). Variants of the attention mechanism that explicitly encode spatial or distance-dependent biases have been previously proposed and successfully applied in different machine-learning settings~\cite{wu2021, sun2022, zhang2025}. Crucially, the inductive bias we propose is Hamiltonian-agnostic: it encodes only a universal geometric bias rather than model-specific information. Empirically, we find that this single ingredient stabilizes optimization on large two-dimensional lattices, starting from random initialization and without providing prior information about the sign structure~\cite{liang2023}, enabling accurate calculations on frustrated systems containing thousands of spins, an ability that has remained out of reach for existing NQS architectures.

To demonstrate the effectiveness of this approach, we apply it to the spin-$\tfrac{1}{2}$ Heisenberg model on the triangular lattice, a paradigmatic example of a frustrated magnet, where geometric constraints prevent the use of sign-problem-free Quantum Monte Carlo methods~\cite{sandvik1999, Sandvik2010, becca2017}. Although there is consensus that the ground state exhibits $120^\circ$ Néel magnetic order, quantitative estimates of the magnetization vary widely $0.16 \lesssim \mathcal{M}_0 \lesssim 0.28$~\cite{Ghioldi2015, Kaneko2014, Heidarian2009, Ghioldi2018, Goetze2016, sorella_sign, White2007, Li2015, Richter2004, Zheng2006, Farnell2014, Yunoki2006, Li2022, huang2024, moss2025rnn, Hasik2024, naumann2025, Li2022} (see \cref{table:magnetization}). This dispersion reflects the difficulty of combining high variational accuracy with controlled finite-size scaling at large system sizes. Our method overcomes this barrier: we reach cluster sizes up to $42\times 42$, obtaining state-of-the-art variational energies and finding the lowest thermodynamic-limit magnetization reported to date, $\mathcal{M}_0 = 0.148(1)$, about $30\%$ of the classical value, highlighting the role of strong quantum fluctuations. Moreover, we show that less accurate variational approximations systematically overestimate the magnetization, providing a concrete explanation for part of the spread in previous results. From finite-size scaling of energy and magnetization, we extract the velocity of low-energy spin excitations and susceptibilities~\cite{batista2016, xie2023, scheie2023} and compare it with linear spin wave theory~\cite{chubukov1999, chernyshev2009, trumper2000} and the Schwinger boson approach~\cite{Ghioldi2018}.

Large-scale access also enables a systematic analysis of the ground-state \emph{sign structure}, a central question in frustrated systems. Using the optimized variational states, we explicitly test candidate sign rules proposed for this model~\cite{huse}, finding that the sign overlap decreases exponentially with system size. This indicates that even extended \emph{local} sign rules rapidly fail to reproduce the exact structure. While this does not constitute a no-go theorem against all possible local basis transformations, it provides quantitative evidence that the sign structure is highly non-local and supports the idea that the Heisenberg antiferromagnet on the triangular lattice has an intrinsic sign problem.

Finally, to demonstrate that our approach is not limited to a particular model, we study the $J_1$-$J_2$ Heisenberg model~\cite{becca2013, nomura2021} on a $20\times 20$ square lattice, achieving state-of-the-art results in comparison to all the other existing variational methods~\cite{chen2024empowering}.

Taken together, these results show that embedding a minimal, physically motivated inductive bias into Transformer-NQS can dramatically improve optimization stability and scalability, thereby enabling controlled access to the thermodynamic limit in frustrated two-dimensional quantum magnets.

\section{Spatial Attention mechanism}\label{sec:att_weights}
Variational approaches based on NQS represent the many-body wave function by mapping spin configurations $\boldsymbol{\sigma} = (\sigma_1, \dots, \sigma_{N})$, with $\sigma_i=\pm 1$, to complex amplitudes $\Psi_{\theta}(\boldsymbol{\sigma})$ parametrized by a neural network with parameters $\theta$. Expectation values are computed using Variational Monte Carlo (VMC)~\cite{becca2017}, and the parameters $\theta$ are optimized via Stochastic Reconfiguration (SR)~\cite{sorella2005, becca2017, rende2024stochastic, chen2024empowering}. 
Among the architectures proposed for frustrated quantum magnets, the Vision Transformer (ViT) wave function~\cite{viteritti2023prl, rende2024fine, viteritti2025prb, Nutakki2025, rende2025foundations, viteritti2025spinglass, chen2025} has demonstrated high accuracy in two dimensions.

The core component of the Transformer architecture is the attention mechanism~\cite{vaswani2023}, which processes a sequence of $n$ input vectors $(\mathbf{x}_1,\dots,\mathbf{x}_n)$, with $\mathbf{x}_i \in \mathbb{R}^d$, and produces an output sequence of equal length $(\mathbf{A}_1,\dots,\mathbf{A}_n)$, referred to as attention vectors, with $\mathbf{A}_i \in \mathbb{R}^d$. A detailed description of how the input vectors $\mathbf{x}_i$ are constructed from a spin configuration $\boldsymbol{\sigma}$ is provided in Sec.~\ref{sec:architecture}. 

In this work, we introduce the \textit{Spatial Attention mechanism} that softly biases the attention weights via a distance-dependent kernel. The resulting attention vectors are given by
\begin{equation}
\label{eq:distance_attn}
\textbf{A}_i = \sum_{j=1}^n \frac{e^{-\gamma d(i,j)}}{\sum_{j'=1}^n e^{-\gamma d(i,j')}} \alpha_{ij} V\textbf{x}_j \ ,
\end{equation}
where the attention coefficients $\alpha_{ij}$ and value matrix $V$ are trainable parameters. The key modification compared to previous attention mechanisms~\cite{viteritti2023prl,viteritti2025prb,rende2024neurips,rende2024prr} is the multiplicative factor depending on the distance $d(i,j)$, defined as the Euclidean distance between lattice sites $i$ and $j$, and the inverse length scale $\gamma>0$, which is learned during optimization and controls the characteristic spatial range of correlations.

Although optimized Transformer wave functions typically exhibit an effective attenuation of attention weights with increasing spatial separation~\cite{rende2025iop,viteritti2025prb,gu2025,rende2024prr}, learning this distance dependence from random initialization becomes progressively more challenging as the system size increases. By explicitly encoding this physically motivated structure, the Spatial Attention mechanism substantially enhances optimization stability in the large-system regime.

Crucially, the proposed modification constitutes a soft and learnable geometric prior that preserves global connectivity, and the architecture continuously reduces to standard attention when the inverse length scales vanish. More flexibility is introduced with the adoption of the Multi-Head Attention mechanism~\cite{vaswani2023}, where each head is associated with an independent length-scale parameter, allowing the Ansatz to represent multiple correlation scales simultaneously~\cite{gu2025}. By employing deep architectures, successive layers can further refine the effective correlation lengths. Refer to \cref{fig:pictorial_representation} for a graphical representation of the mechanism.

Operationally, \cref{eq:distance_attn} constitutes a simple modification of the standard Attention mechanism and can be implemented as a one-line change in the  ViT architecture~\footnote{See the NetKet tutorial \url{[https://netket.readthedocs.io/en/latest/tutorials/ViT-wave-function.html}.}.\\

\begin{table}[t]
\centering
\begin{tabular}{c c c c}
\hline\hline
\textbf{Size} & \textbf{Ansatz} & \textbf{Energy} & \textbf{Ref.} \\
\hline
\multirow{6}{*}{$18 \times 18$}
 & p-BCS & $-0.53570(1)$ & ~\cite{sorellaprb2006} \\
 & Jastrow-Gutzwiller & $-0.54542(1)$ & ~\cite{beccaprb2016} \\
 & EPS & $-0.5459(1)$ & ~\cite{mezzacapo2010iop} \\
 & RVB & $-0.54716(3)$ & ~\cite{beccaprb2009} \\
 & RNN & $-0.54861(1)$ & ~\cite{moss2025rnn} \\
 & \textbf{ViT} & \textbf{-0.551703(1)} & \textbf{This work} \\
 & \textbf{Zero Variance} & \textbf{-0.55200(1)} & \textbf{This work} \\
 \hline
\multirow{3}{*}{$24 \times 24$}
 & RNN & $-0.54810(1)$ &~\cite{moss2025rnn} \\
 & \textbf{ViT} & \textbf{-0.551602(1)} & \textbf{This work} \\
 & \textbf{Zero Variance} & \textbf{-0.55189(1)} & \textbf{This work} \\
\hline
\multirow{3}{*}{$30 \times 30$}
 &  Jastrow-Gutzwiller& $-0.545348$ & ~\cite{beccaprb2016} \\
 & RNN & $-0.54793(1)$ & ~\cite{moss2025rnn} \\
 & \textbf{ViT} & \textbf{-0.551528(1)} & \textbf{This work} \\
& \textbf{Zero Variance} & \textbf{-0.55180(1)} & \textbf{This work} \\

\hline
\multirow{1}{*}{$42 \times 42$}
 & \textbf{ViT} & \textbf{-0.551429(1)} & \textbf{This work} \\
 & \textbf{Zero Variance} & \textbf{-0.55171(1)} & \textbf{This work} \\
\hline\hline
\end{tabular}
\caption{\label{table:energies} \textbf{Variational energies.} Ground-state energies per site for various Ansatze and system sizes from $L=18$ to $L=42$. The results of this work are obtained by enforcing translational and $C_{6v}$ point-group symmetries (refer to \cref{sec:architecture}). For each system size the zero-variance extrapolated energy is also reported for comparison (refer to \cref{sec:en_var}).}
\end{table}

\begin{figure*}[ht]
    \begin{center}
\centerline{\includegraphics[width=1.8\columnwidth]{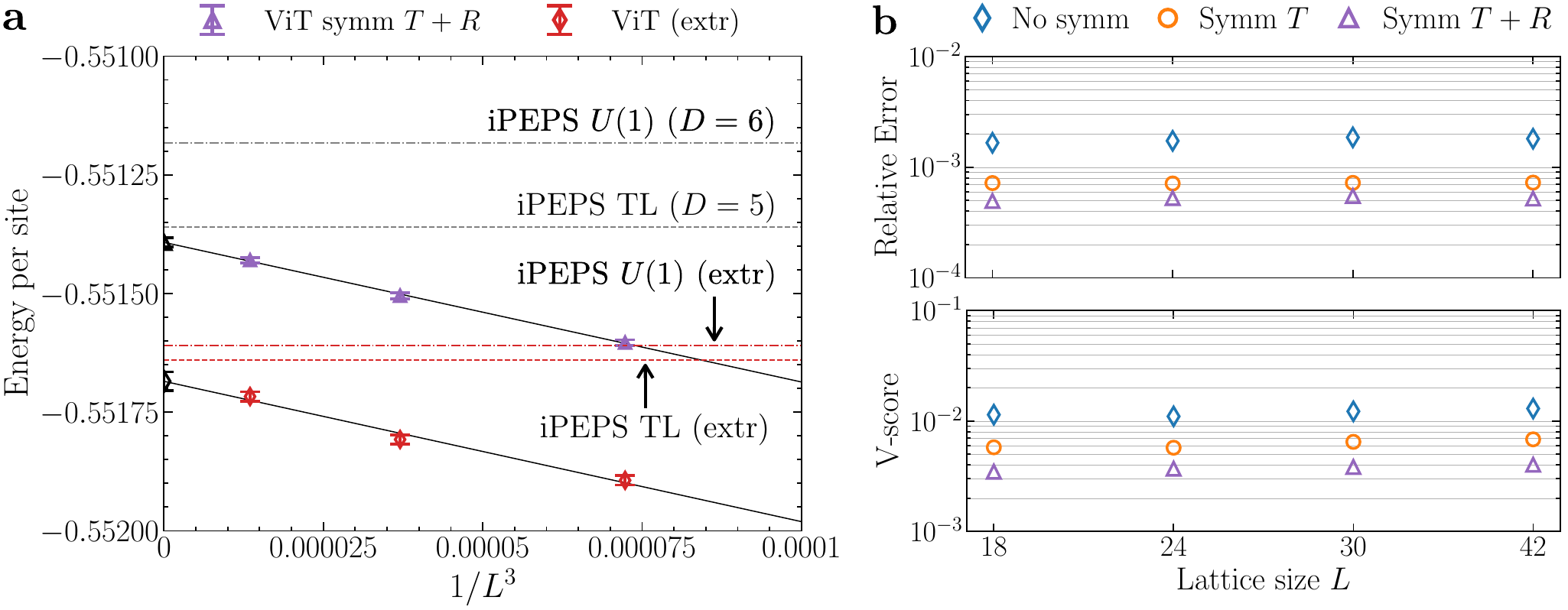}}
        \caption{\label{fig:energy}\textbf{Energy scaling and size-consistent accuracy.} \textbf{a.} Finite-size extrapolations of the ground-state variational energy per site as a function of $1/L^{3}$, using data from lattice sizes $L = 24, 30,$ and $42$. Results obtained with the fully symmetrized ViT wave function are shown as violet triangles, while red rhombi denote energies obtained from zero-variance extrapolations (see Sec.~\ref{sec:en_var} for details). These results are compared with the most recent and accurate thermodynamic-limit estimates from iPEPS calculations, including iPEPS TL~\cite{naumann2025} and 
        iPEPS $U(1)$~\cite{Hasik2024}. Extrapolated energies are not shown for methods whose best variational energy exceeds $-0.551$. \textbf{b.} Top panel: relative error of the variational energies with respect to the zero-variance extrapolated values as a function of system size $L$, for different symmetry projections: no symmetry projection (No symm, blue rhombi), translation symmetry projection (Symm $T$, orange circles), and full translation and rotation symmetry projection (Symm $T+R$, violet triangles). Bottom panel: corresponding V-score values.}
    \end{center}
\end{figure*}
\section{Heisenberg Triangular lattice}
As a concrete and physically relevant application of our approach, we investigate the ground state properties of the spin-$\tfrac{1}{2}$ Heisenberg Hamiltonian on the triangular lattice
\begin{equation}
    \hat{H} = J\sum_{\langle i,j \rangle} \hat{\boldsymbol{S}}_i \cdot \hat{\boldsymbol{S}}_j \ ,
\end{equation}
where $\hat{\boldsymbol{S}}_{i}=(S_i^x,S_i^y,S_i^z)$ is the spin-$1/2$ operator at site $i$, and $J > 0$ is the antiferromagnetic coupling among nearest-neighbors. All calculations are performed on periodic $L \times L$ clusters with $L$ chosen as a multiple of $3$, ensuring commensurability with the three-sublattice $120^\circ$ N\'eel order~\cite{sorella_sign, chernyshev2007}. 
\subsection{Energy and Magnetization}\label{sec:energy_and_magn}
To assess the quantitative performance of the spatial-attention–based ViT wave function, we first benchmark it against leading variational Ans\"atze. As summarized in Table~\ref{table:energies}, the ViT achieves significantly lower energies than Gutzwiller-projected states~\cite{sorellaprb2006, beccaprb2016} and recent neural-network wave functions based on Recurrent Neural Networks (RNN)~\cite{rnn} across all accessible system sizes. Notably, on the largest cluster studied ($42\times 42$) the variational energy is below the best previously reported value for the $18\times 18$ cluster. Overall, the ViT improves the relative energy accuracy by more than an order of magnitude compared to other approaches.

We next exploit the large accessible sizes to perform controlled finite-size scaling of the energy. For a two dimensional antiferromagnet, the leading correction to the thermodynamic-limit energy is expected to scale as $1/L^{3}$~\cite{Fisher1989, sandvik1997, chubukov1999, sandvik2026} (refer to \cref{sec:sign_magn} and \textit{Supplementary Information} for additional details). As shown in \cref{fig:energy}$\textbf{a}$, the numerical data of the variational energy per site follow a linear scaling in $1/L^3$ for $L \ge 24$ (purple triangles), allowing for a reliable extrapolation to $L\to +\infty$. We compare our extrapolated results with the most accurate tensor-network estimates obtained from iPEPS calculations performed directly in the thermodynamic limit at finite bond dimension $D$ (black horizontal dashed lines)~\cite{Hasik2024,naumann2025}. Remarkably, our results are below the iPEPS estimations up to $L=42$ and reach state-of-the-art results in the thermodynamic limit.
Moreover, for each system size, we perform an energy-variance extrapolation (see \cref{sec:en_var}) and compare the extrapolated energies (red rhombi) with the iPEPS extrapolations obtained in the limit $D\rightarrow +\infty$ (red horizontal dashed lines). From this analysis, we estimate a thermodynamic-limit energy per site of -0.55168(2), which is lower than the corresponding iPEPS estimates.

In \cref{fig:energy}\textbf{b} we report estimations of the accuracy of the variational state.
In the upper panel we show the relative energy error as a function of the system size for three
levels of symmetry restoration (none, translations only, full point-group $C_{6v}$), computed with respect to the corresponding
zero-variance extrapolations (see \cref{sec:en_var}). In the lower panel of the same figure, we also present the V-score~\cite{vscore}, a reference-free accuracy metric for variational methods. Both measures remain essentially stable as the system size grows. Importantly, this is achieved with an approximately fixed parameter count,
$P\approx 4.5\times 10^{5}$, for all $L$ (see \cref{sec:architecture}), indicating that accuracy does not rely on scaling the network with the lattice size. \emph{Size consistency} is a key requirement for a scalable variational Ansatz, and in practice, many variational states achieve high accuracy on small lattices but exhibit a systematic degradation in performance as the cluster size grows~\cite{viteritti2022}.

\begin{figure*}[ht]
    \begin{center}
\centerline{\includegraphics[width=1.8\columnwidth]{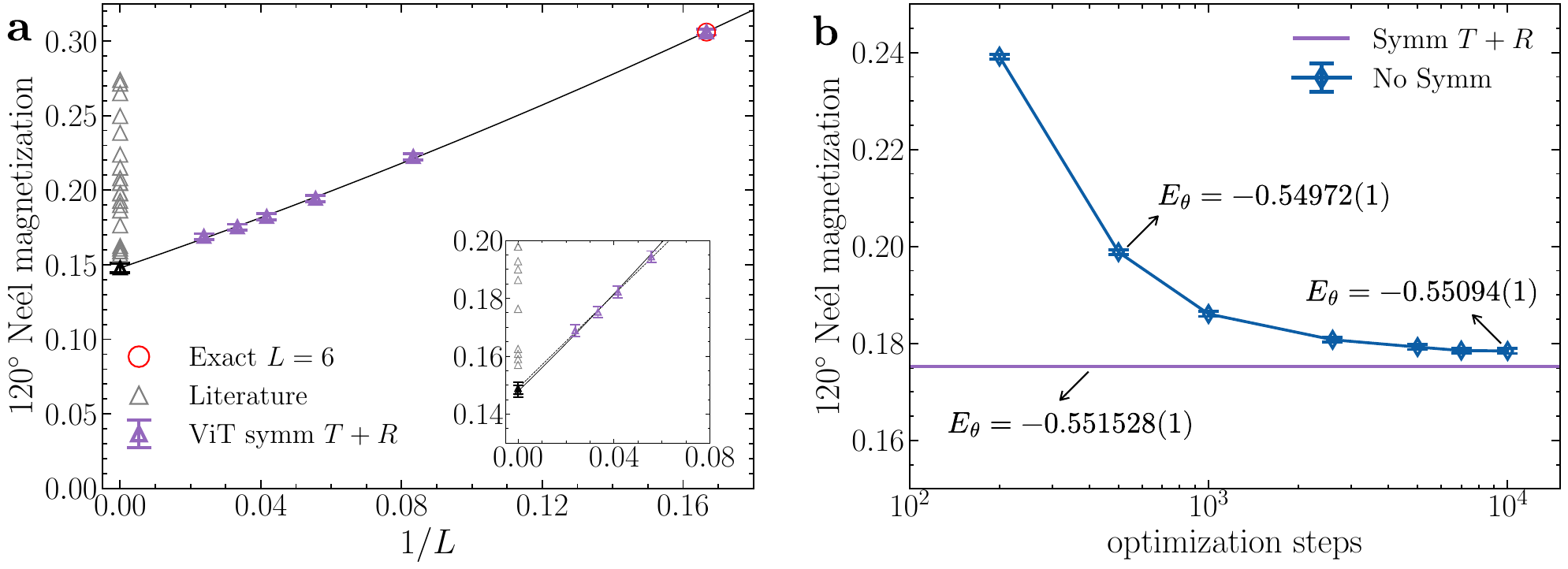}}
        \caption{\label{fig:magnetization}\textbf{Size-scaling of the magnetization.} \textbf{a}. Finite-size extrapolation of the magnetization order parameter, as defined in in the main text, to the thermodynamic limit (violet triangles). All the available results in literature are showed for comparison (gray triangles) and reported in \cref{table:magnetization} in \cref{sec:magn_lit}. 
        \textbf{Inset:} Zoom of the same quantity for $1/L \rightarrow 0$ in which different extrapolations are shown, quadratic (solid line) and linear (dashed line).  \textbf{b}. Magnetization order parameter as a function of optimization steps for the $L=30$ cluster, comparing the fully symmetrized ViT (violet horizontal line) with the non-symmetrized ViT (blue rhombi). Reference values of the ground-state variational energies at selected optimization steps and the final energy for the fully symmetrized state are also shown.}
    \end{center}
\end{figure*}

Let us now move to the study of the magnetic properties,
assessed through the static spin structure factor ${C(\boldsymbol{q}) = \sum_{\boldsymbol{R}} e^{i \boldsymbol{q} \cdot \boldsymbol{R}} \langle \hat{\boldsymbol{S}}_{\boldsymbol{0}} \cdot \hat{\boldsymbol{S}}_{\boldsymbol{R}} \rangle}$, where the expectation value is evaluated over the variational state.
For a periodic cluster $L\times L$, the order parameter of the $120^{\circ}$ Néel order is detected by ${\mathcal{M}^2(L) = C(\boldsymbol{Q})/L^2}$ with ${\boldsymbol{Q}=(4\pi/3,0)}$~\cite{beccaprb2009, moss2025rnn}. In \cref{fig:magnetization}\textbf{a} we show the order parameter $\mathcal{M}(L)$ versus $1/L$ and its extrapolation to the thermodynamic limit. Our data yield a thermodynamic-limit magnetization $\mathcal{M}_{0}=0.148(1)$, the smallest among all the available estimates (see \cref{table:magnetization} for a collection of previous calculations). In particular, we emphasize that the magnetization predicted by linear spin-wave theory (SW) is 0.239~\cite{chubukov1999,chernyshev2009,trumper2000},
exhibiting a relative error of approximately $62\%$ with respect to the fully quantum
result. This differs significantly from the square-lattice Heisenberg model, where
SW predicts the magnetization with an error of only about $1\%$~\cite{igarashi1992,sandvik1997}.
This comparison suggests that, despite the presence of magnetic order in the
triangular-lattice Heisenberg model, achieving a quantitatively accurate fully
quantum description remains highly nontrivial. 
To clarify why magnetization estimates can vary widely across methods (refer to \cref{table:magnetization}), in \cref{fig:magnetization}\textbf{b} we track the order parameter during the optimization procedure for $L=30$. 
As the energy decreases and the wave function becomes more accurate, the magnetization monotonically reduces~\cite{viteritti2025prb}. This
directly indicates that limited variational accuracy can lead to a systematic \emph{overestimate} of magnetic order and highlights the importance of high-precision wave-function optimization in frustrated systems.

\subsection{Finite-Size Scaling}
Tensor-network methods such as iPEPS operate directly in the thermodynamic limit and have achieved remarkable success in describing ground-state energies and local observables~\cite{Vlaar2021, Corboz2012, Corboz2016, Jordan2008, verstraete2004, Hasik2024,naumann2025}. At the same time, working exclusively in the infinite-system setting comes with intrinsic limitations: quantities that rely on finite-size scaling cannot be accessed. By contrast, accurate variational wave functions defined on large periodic lattices allow for controlled finite-size scaling analyses of both the energy and order parameters, providing direct access to physically relevant properties. In the case of the antiferromagnetic triangular lattice, analytic expressions for the leading finite-size corrections can be derived and used to quantitatively fit the numerical data.

\begin{figure*}[ht]
    \begin{center}
\centerline{\includegraphics[width=1.8\columnwidth]{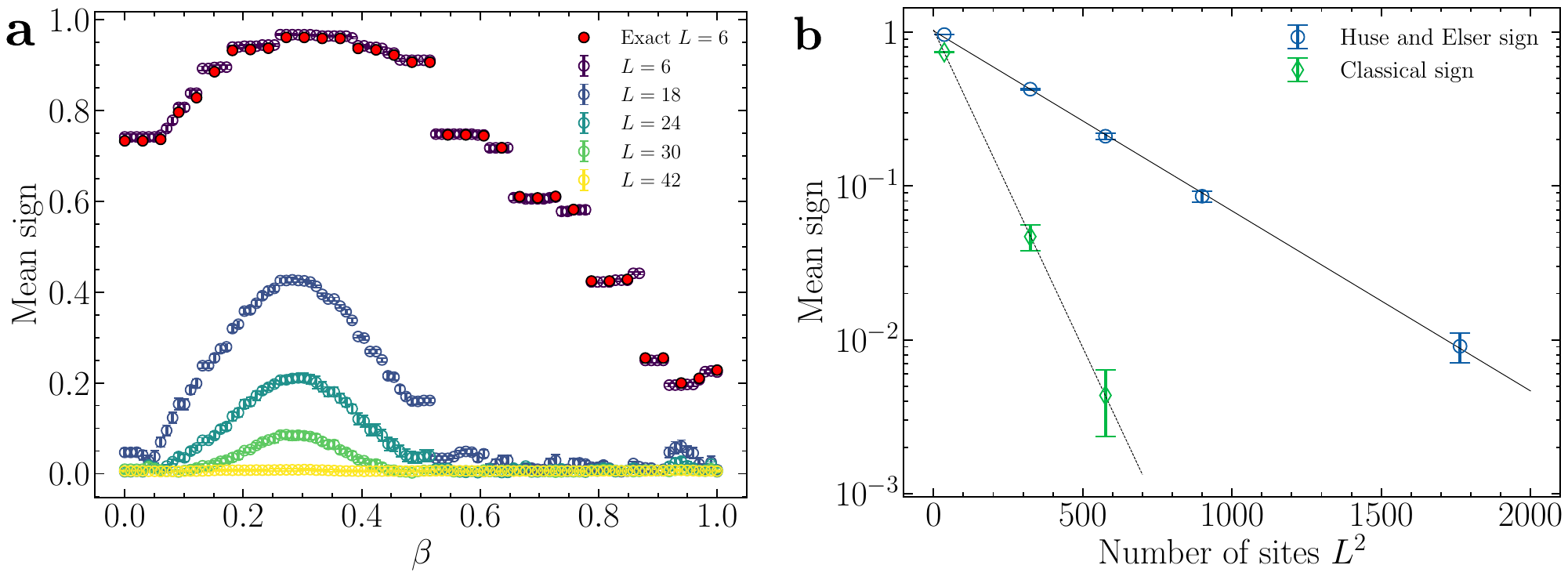}}
        \caption{\label{fig:sign_structure}\textbf{Ground-state sign structure.} \textbf{a.} Mean sign overlap [see definition in \cref{eq:mean_sign}] between the ViT state and the Huse–Elser prescription~\cite{huse, Capriotti1999} as a function of its variational parameter $\beta$. Exact results are shown for $L = 6$, for which the sign overlap between the ViT wave function and the exact ground state is ${\langle s \rangle = 0.99999707}$. \textbf{b.} Mean sign overlap as a function of the number of sites $L^2$ between the Transformer wave function and the classical sign structure (green circles) or the Huse–Elser wave function~\cite{huse} at the optimal $\beta \approx 0.27$ (blue circles).}
    \end{center}
\end{figure*}

At low energy and long distances, the spin fluctuations on the triangular lattice can be captured by the parameterization
\begin{equation}
{\boldsymbol S}_i (\tau) = \mathcal{M}_0 \left[ {\boldsymbol n}_1 ({\boldsymbol x}_i , \tau) \cos ({\boldsymbol Q} \cdot {\boldsymbol x}_i) + {\boldsymbol n}_2 ({\boldsymbol x}_i , \tau) \sin ({\boldsymbol Q} \cdot {\boldsymbol x}_i) \right]\ ,
\end{equation}
where $\mathcal{M}_0$ is the ground state ordered moment in the infinite lattice, ${\boldsymbol Q} = (4 \pi/3, 0)$
 is the ordering wavevector, and 
${\boldsymbol n}_{1},{\boldsymbol n}_{2}$ are slowly varying functions of space (${\boldsymbol x}$) and imaginary time ($\tau$) obeying the constraints $|{\boldsymbol n}|_1^2 = |{\boldsymbol n}_2|^2 = 1$, ${\boldsymbol n}_1 \cdot {\boldsymbol n}_2 = 0$.  It is useful to solve these constraints in terms of unit length complex spinor $ z \equiv (z_1, z_2)$, $|z_1|^2 + |z_2|^2= 1$,  by the parameterization ${n_{1a} + i n_{2a} = z_\gamma \varepsilon_{\gamma\alpha}  \sigma^a_{\alpha\beta} z_\beta}$
where $\alpha,\beta=1,2$, $\varepsilon$ is the unit antisymmetric matrix, $a=x,y,z$, and $\sigma^a$ are the Pauli matrices~\cite{Angelucci1991,Chubukov1994,Chubukov1994a}. From the symmetry of the ordered state, we obtain the low energy description as a non-linear $\sigma$-model
\cite{Chubukov1994,Chubukov1994a} 
\begin{equation}
{\cal S} =
\int d^2 x d \tau \, \sum_{\mu = {\boldsymbol x}, \tau} \frac{1}{g_{\mu}}
\left[ \partial_{\mu} z^{\dagger} \partial_{\mu} z
- \frac{\gamma_{\mu}}{4}
\left( z^{\dagger} \partial_{\mu} z - \partial_{\mu}
z^{\dagger} z \right)^2 \right] \ .
\label{eq:action_triangular}
\end{equation}
For small fluctuations about the ordered state, we solve the unit length constraint by writing ${z_1 = \sqrt{1 - \pi_1^2 - \pi_2^2 - \pi_3^2} + i \pi_1}$ and $z_2 = \pi_2 + i \pi_3$ where $\pi_{\ell}$ ($\ell=1,2,3$) are three real fields.
Inserting this in \cref{eq:action_triangular}, and expanding to quadratic order, we obtain a harmonic theory of three spin wave modes
\begin{equation}
{\cal S}_{\text{SW}} = \int d^2 x d \tau \sum_{\ell = 1}^3 \left[ \frac{m_\ell}{2} \left\{ (\partial_\tau \pi_\ell)^2 + c_\ell^2 (\nabla \pi_\ell)^2 \right\} 
\right].
 \end{equation}
The couplings are defined in terms of two spin-wave velocities $c_1 = c_\parallel$, $c_2 = c_3 = c_\perp$ and the uniform magnetic susceptibilities
 \begin{equation}
     \chi_\perp = \tfrac{1}{2 g_\tau} = \tfrac{m_2}{4} = \tfrac{m_3}{4}  \quad \quad  \chi_\parallel = \tfrac{1}{2 g_\tau} (1 + \gamma_\tau)  = \tfrac{m_1}{4} \ .
 \end{equation}
As shown in the \textit{Supplementary Information}, computations from the spin-wave theory $\mathcal{S}_{\text{SW}}$ lead to the following results for the finite-size corrections to the ground state energy and magnetization squared (measured from the equal-time structure factor)
\begin{align}
    \frac{E}{L^2} &= e_0 + \left( \sum_{\ell=1}^{3} \frac{c_\ell }{2} \right) \frac{\mathcal{C}_1} {L^{3}} + \dots
    \label{eq:scaling_energy} \\
    \mathcal{M}^2(L) &= \mathcal{M}^2_0\left[ 1 + \left(\sum_{\ell = \parallel,\perp} \frac{1}{2 c_\ell\chi_\ell} \right) \frac{\mathcal{C}_2}{L} \right] + \dots
    \label{eq:scaling_magn}
\end{align}
Here, $L$ is the linear size of the lattice, $e_0$ and $\mathcal{M}_0$ are the ground state energy per site and ordered moment in the infinite lattice, respectively. The constants $\mathcal{C}_{1}$ and $\mathcal{C}_{2}$ are universal numbers depending only on the large scale geometry of the lattice and are computed explicitly in the \textit{Supplementary Information} \cite{Latticesums}.

Fitting our numerical large-$L$ data with Eqs.~\eqref{eq:scaling_energy} and \eqref{eq:scaling_magn} we obtain $\bar{c}\approx 3.88$ and $\overline{c\chi} \approx 0.061$, where ${\bar{c}=c_{\parallel} + 2c_{\perp}}$ and $\overline{c\chi} = \left(\tfrac{1}{2\chi_\parallel c_{\parallel}}+\tfrac{1}{2\chi_\perp c_{\perp}}\right)^{-1}$. These values differ from the linear spin-wave values $\bar{c}_{\mathrm{SW}}\approx 3.14$, $\overline{c\chi}_{\text{SW}} \approx 0.086$~\cite{chubukov1999,chernyshev2009} (see \textit{Supplementary Information} for additional details) and Schwinger boson ${\bar{c}_{\mathrm{SB}}\approx 3.15}$~\cite{Ghioldi2018} estimates. 
Together with the strong renormalization
of the magnetization $\mathcal{M}_0$ (refer to \cref{sec:energy_and_magn} and \cref{table:magnetization}), this highlights that a quantitatively reliable description of the triangular-lattice Heisenberg
antiferromagnet requires the inclusion of strong quantum effects that are not captured by SW contrary to the square lattice case. Importantly, spin-wave velocities and susceptibilities are experimentally accessible quantities, and the values reported here therefore provide a stringent benchmark for candidate materials realizing the triangular-lattice Heisenberg model~\cite{Susuki2013, xie2023, scheie2023, Scheie2024}.

\subsection{Non-Local Sign Structure}\label{sec:sign_magn}
Beyond energies and order parameters, frustrated magnets raise a fundamental question about the structure of the ground state: \textit{is the sign problem a removable basis artefact, or does it reflect a genuinely non-local sign structure?} For the pure Heisenberg model on bipartite lattices, the Marshall-Peierls sign rule provides a local basis transformation that renders the ground state non-negative in the computational basis~\cite{marshall1955}. No analogous prescription is known for the triangular lattice.

\begin{figure}[t]
    \begin{center}
\centerline{\includegraphics[width=0.9\columnwidth]{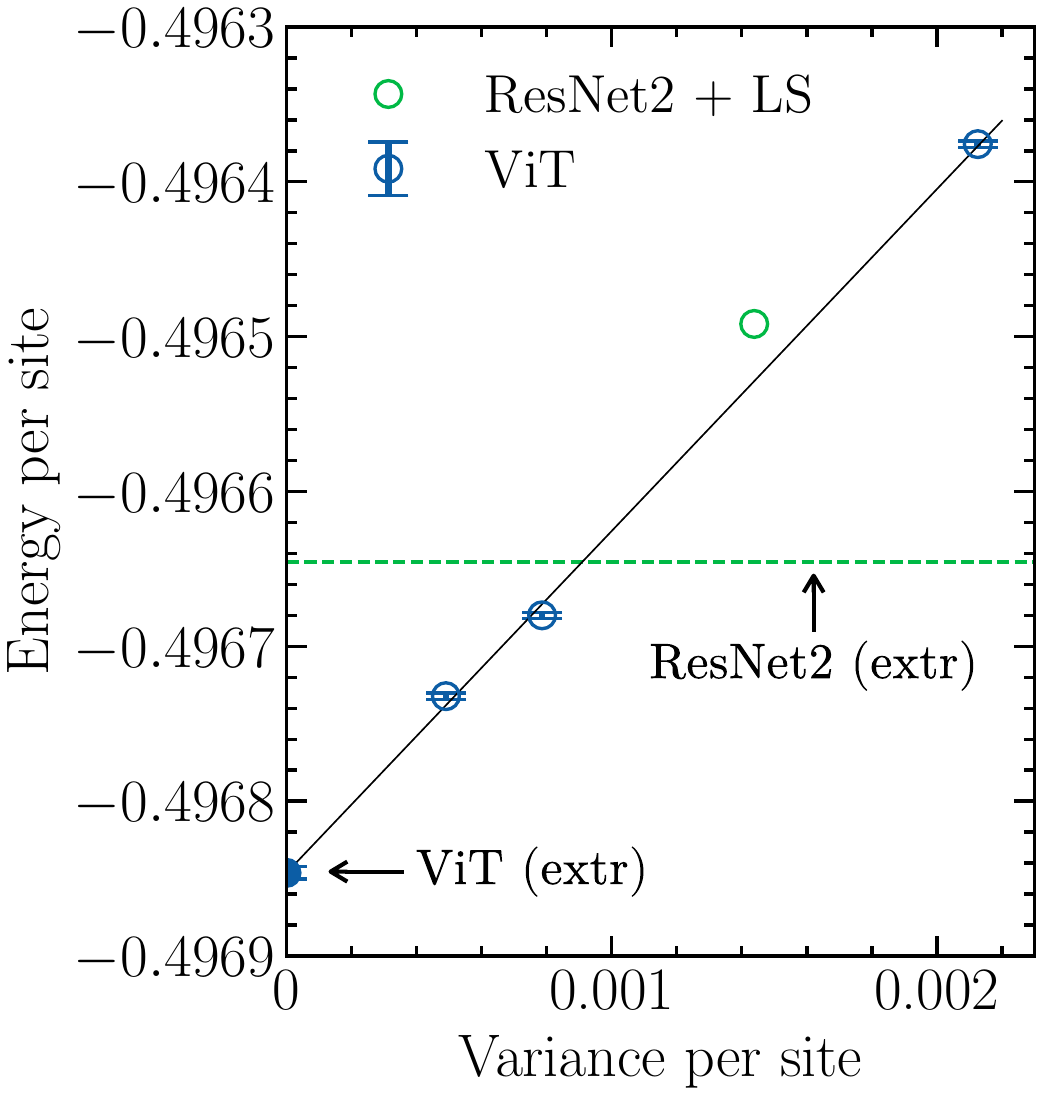}}
        \caption{\label{fig:J1J2_square} \textbf{Energy $J_1$-$J_2$ Heisenberg model.} Variational energies for $J_1$-$J_2$
 Heisenberg model at $J_2/J_1=0.5$ on a $20\times 20$ cluster with periodic boundary conditions plotted as a function of the variance per site. The three blue empty points correspond to ViT wave functions with: no symmetry, translational symmetry, and full point-group symmetry restoration. A linear extrapolation of these points to zero variance yields an energy of $-0.49684(1)$ (blue full circle). The green circle show the variational energy obtained with the ResNet$2$ Ansatz with one Lanczos step, together with the zero-variance extrapolated value (see dashed green line) reported in Ref.~\cite{chen2024empowering}.}
 \end{center}
\end{figure}

A notable attempt in this direction was proposed by~\citet{huse}, who introduced an extended sign construction combining the classical three-sublattice phase factor
\begin{equation}
    {T_{\text{class}}(\boldsymbol{\sigma})=\exp\!\left[i \frac{2\pi}{6}\left( \sum_{i \in \mathrm{B}} \sigma_i^z -  \sum_{i \in \mathrm{C}} \sigma_i^z \right)\right]} \ ,
\end{equation}
with $B$ and $C$ indicating two of the three sublattices of the triangular lattice, along with an additional three-body contribution 
\begin{equation}
    {T_{3\text{body}}(\boldsymbol{\sigma}) = \exp\!\left(i \frac{\beta}{8} \sum_{\langle i,j,k \rangle} \gamma_{ijk} \sigma_i^z \sigma_j^z \sigma_k^z\right)} \ ,
\end{equation}
leading to the sign prescription ${\Phi_{\mathrm{HE}}(\boldsymbol{\boldsymbol{\sigma}}) = \Re\{T_{3\text{body}}(\boldsymbol{\sigma})T_{\text{class}}(\boldsymbol{\sigma})\}}$~\cite{huse,sorella_sign}.
Here, $\beta$ is the only variational parameter, and the coefficients $\gamma_{ijk} = 0, \pm 1$ are chosen to preserve the symmetries of the classical Néel state~\cite{huse, sorella_sign}.

To quantify the agreement between sign structures, we follow Ref.~\cite{sorella_sign} and define the mean sign overlap for a variational state $\Psi_{\theta}(\boldsymbol{\sigma})$ as
\begin{equation}\label{eq:mean_sign}
    \langle s \rangle = \sum_{\boldsymbol{\sigma}} |\Psi_{\theta}(\boldsymbol{\sigma})|^{2} \, \mathrm{sgn}\!\left\{ \Phi_{\text{HE}}(\boldsymbol{\sigma}) \Re [\Psi_{\theta}(\boldsymbol{\sigma})]\right\}.
\end{equation}
which can be stochastically estimated by sampling from $|\Psi_{\theta}(\boldsymbol{\sigma})|^{2}$. In our case, the use of the real part of $\Psi_{\theta}(\boldsymbol{\sigma})$ is justified by the fact that the phase
distribution is sharply peaked around $0$ and $\pi$ for the optimized states (see \cref{sec:phases_distrib}).
In \cref{fig:sign_structure}\textbf{a}, we show the mean sign overlap between the ViT and the Huse–Elser state as a function of $\beta$. Moreover, for $L=6$ we also compare with the exact ground state; the ViT achieves a nearly perfect sign overlap, $\langle s \rangle = 0.99999707$, confirming that the architecture can represent the correct sign structure at small sizes. While for $L=6$ the Huse–Elser sign rule matches the exact signs well (for $\beta \approx 0.27$, $\langle s \rangle \approx 0.96$)~\cite{Capriotti1999}, the agreement deteriorates rapidly as the system grows. In \cref{fig:sign_structure}\textbf{b} , we measure the mean sign overlap as a function of the number of sites $L^2$; for each size, we select the optimal $\beta$ for which the overlap is maximum. The overlap decays exponentially with the system size. The learned sign structure, therefore, differs markedly from both classical and the prescription proposed by~\citet{huse}, indicating that sign information not captured by extended local prescriptions becomes increasingly important at large sizes. Interestingly, we observe distinct decay rates for the classical ($\beta=0$) and Huse–Elser sign rules. This behavior suggests that systematically incorporating higher-order many-body contributions can further improve the overlap. At the same time, it indicates that the sign structure of the triangular lattice may be intrinsically non-local, such that no local unitary rotation in the computational basis can render the wave function positive definite~\cite{torlai2020}.

\section{$J_1$-$J_2$ Heisenberg Square Lattice }\label{sec:square_lattice}
The improvement provided by the Spatial Attention mechanism is not specific to triangular geometries or to Hamiltonians with only nearest-neighbor interactions. Because it relies solely on geometric information, the underlying physical principles are expected to remain effective across a broad class of models.

To demonstrate this generality, we consider the spin-$1/2$ $J_1$-$J_2$ Heisenberg model on a $20\times20$ square lattice with periodic boundary conditions,
\begin{equation}
    \hat{H} =
    J_1\sum_{\langle i,j\rangle} \hat{\boldsymbol{S}}_i \cdot \hat{\boldsymbol{S}}_j
    +
    J_2\sum_{\langle\!\langle i,j\rangle\!\rangle} \hat{\boldsymbol{S}}_i \cdot \hat{\boldsymbol{S}}_j ,
\end{equation}
at the highly frustrated point $J_2/J_1 = 0.5$, where various numerical studies have proposed the existence of a quantum spin liquid~\cite{nomura2021,becca2013}. The competing interactions in the strongly frustrated regime make the model a stringent benchmark for any variational Ansatz.

We compute the variational energy and obtain state-of-the-art results that surpass those achieved by a deep Residual Convolutional Neural Network augmented with a Lanczos step (ResNet2) from Ref.~\cite{chen2024empowering}. In \cref{fig:J1J2_square}, we report the energies obtained with increasing levels of symmetry restoration: no symmetry [$-0.496375(1)$], translational symmetry [$-0.496698(1)$], and full point-group ($C_{4v}$) symmetry [$-0.496732(1)$]. A zero-variance extrapolation yields an energy of $-0.49684(1)$ (see Sec.~\ref{sec:en_var} for details)~\cite{becca2015,becca2013}. The fully symmetric state attains a V-score of approximately $0.0019$~\cite{vscore} and a relative error $\Delta\varepsilon \approx 2.3\times10^{-4}$ with respect to the extrapolated value, demonstrating high accuracy even in this strongly frustrated regime.

Notably, the ViT Ansatz with translational symmetry alone already achieves an energy lower than both the ResNet2 result with one Lanczos step~\cite{becca2015} and its zero-variance-extrapolated estimate reported in Ref.~\cite{chen2024empowering}. This further demonstrates that the proposed approach is not limited to the triangular-lattice Heisenberg model but extends naturally to other challenging quantum many-body Hamiltonians.

\section{Discussion} 
This work establishes a concrete route for extending variational methods to lattice sizes that were previously out of reach. We introduced the Spatial Attention mechanism, a minimal geometric inductive bias within the Vision Transformer architecture that fundamentally changes the practical regime accessible with Neural-Network Quantum States: it stabilizes large-scale optimization to the extent that controlled finite-size scaling becomes feasible even in frustrated two-dimensional systems.
For the triangular-lattice Heisenberg antiferromagnet, this capability enables highly accurate estimates of thermodynamic-limit quantities, such as the ground-state energy, magnetization, and, in turn, experimentally measurable parameters such as spin-wave velocities and susceptibilities. It also clarifies a systematic source of variability in earlier estimates of the order parameter: less accurate variational states tend to overestimate magnetic order, leading to an apparent enhancement of magnetization.

Our results indicate that further extensions to even larger system sizes are primarily constrained by available computational resources rather than by conceptual or methodological limitations. Looking ahead, continued advances in hardware are expected to substantially increase the available computational capacity. Moreover, the inherently parallel structure of the Variational Monte Carlo framework makes the present approach well suited for efficient execution on large GPU clusters, enabling simulations to scale to larger lattice sizes at comparable accuracy without requiring fundamental modifications to the methodology. Promising future directions include improving optimization efficiency~\cite{shokry2025}, enhancing sampling strategies~\cite{misery2025}, and reducing the computational cost of observable estimation through the use of mixed precision algorithms~\cite{solinas2026} and controlled approximations~\cite{clark2025prb} that enable scaling to increasingly large systems.

We expect our variational approach to become a leading tool for the discovery and characterization of exotic phases of matter, such as quantum spin liquids~\cite{Balents2010, savary2016}. Beyond spin models, a natural extension is the use of this architecture as a backflow Ansatz for fermionic wave functions~\cite{clark2025prb, robledomoreno2022, roth2025}. In this context, access to very large system sizes is essential to address long-standing and debated questions, such as the emergence of superconductivity in the two-dimensional Hubbard model~\cite{qin2020, xu2024, roth2025}, by enabling the direct evaluation of pairing correlation functions at increasingly large distances. Finally, the bias we introduce is entirely Hamiltonian-agnostic, providing a flexible foundation for future developments that extend variational simulations to increasingly complex Hamiltonians, well beyond the models considered here.

\section*{Methods}

\subsection{Literature Survey of Magnetization Estimates}\label{sec:magn_lit}
In this Section, we compile reference values for the ground-state magnetization in the thermodynamic limit as reported in the literature using a variety of numerical and analytical methods. These estimates are summarized in \cref{table:magnetization} and ordered from larger to smaller magnetization. Notably, our result yields the smallest magnetization among all values currently available in the literature. Moreover, among approaches based on finite-size calculations, our study is performed on the largest cluster size considered to date.

\begin{table}[t]
\centering
\begin{tabular}{lcccc}
\hline\hline
\textbf{Method} & \textbf{Magnetization} & \textbf{Max.~Size} & \textbf{Year} & \textbf{Ref.} \\
\hline
SB      & 0.2739      & TL              & 2015 & \cite{Ghioldi2015}    \\
VMC     & 0.2715(30)  & 324             & 2014 & \cite{Kaneko2014}     \\
VMC     & 0.265       & 576             & 2009 & \cite{Heidarian2009}  \\
SWT     & 0.24974     & TL              & 2009 & \cite{Ghioldi2015}    \\
SWT     & 0.2386      & TL              & 2015 & \cite{Ghioldi2015}    \\
iPEPS   &0.2327       & TL              & 2022 & \cite{chi2022}        \\
SB+1/N  & 0.224       & 432             & 2018 & \cite{Ghioldi2018}    \\
CC      & 0.21535     & $10^*$          & 2016 & \cite{Goetze2016}     \\
DMRG    & 0.208(8)    & 312             & 2024 & \cite{huang2024}      \\
GFMC    & 0.205(10)   & 144             & 1999 & \cite{sorella_sign}   \\
DMRG    & 0.205(15)   & $\approx140$    & 2007 & \cite{White2007}      \\
RNN     & 0.198(2)    & 900             & 2025 & \cite{moss2025rnn}    \\
CC      & 0.198(5)    & $10^*$          & 2015 & \cite{Li2015}         \\
SE      & 0.198(34)   & TL              & 2015 & \cite{Zheng2006}      \\
ED      & 0.193       & 36              & 2004 & \cite{Richter2004}    \\
SE      & 0.19(2)     & TL              & 2006 & \cite{Zheng2006}      \\
CC      & 0.1865      & $10^*$          & 2014 & \cite{Farnell2014}    \\
FNE     & 0.1765(35)  & 324             & 2006 & \cite{Yunoki2006}     \\
FN      & 0.1625(30)  & 324             & 2006 & \cite{Yunoki2006}     \\
PESS    & 0.161(5)    & TL              & 2022 & \cite{Li2022}         \\
iPEPS   & 0.159(6)    & TL              & 2025 & \cite{Hasik2024}      \\
iPEPS   & 0.157       & TL              & 2025 & \cite{naumann2025}    \\
\hline
\textbf{ViT} & \textbf{0.148(1)} & \textbf{1764} & \textbf{2025} & \textbf{This work} \\
\hline\hline
\end{tabular}
\caption{\label{table:magnetization} \textbf{Table of magnetizations.}
Ground-state magnetization in the thermodynamic limit for the spin-$\tfrac{1}{2}$ Heisenberg model on the triangular lattice, obtained using various approaches. SB, CC, SWT, VMC, FN, and FNE denote Schwinger-boson mean-field theory, coupled-cluster method, spin-wave theory, variational Monte Carlo, fixed-node Monte Carlo, and fixed-node Monte Carlo with an effective Hamiltonian, respectively. For the three CC studies~\cite{Goetze2016,Li2015,Farnell2014}, the maximal
parameters $n$ used in the lattice-animal-based subsystem (LSUB$n$)
scheme are listed and indicated by superscript ($*$). For all other methods, the maximum system size (number of lattice sites) used in the calculations is listed. Approaches that operate directly in the thermodynamic limit are labeled by TL. The table is adapted from Ref.~\cite{Li2022} and updated to include the most recent results.}
\end{table} 

\begin{figure}[t]
    \begin{center}
\centerline{\includegraphics[width=\columnwidth]{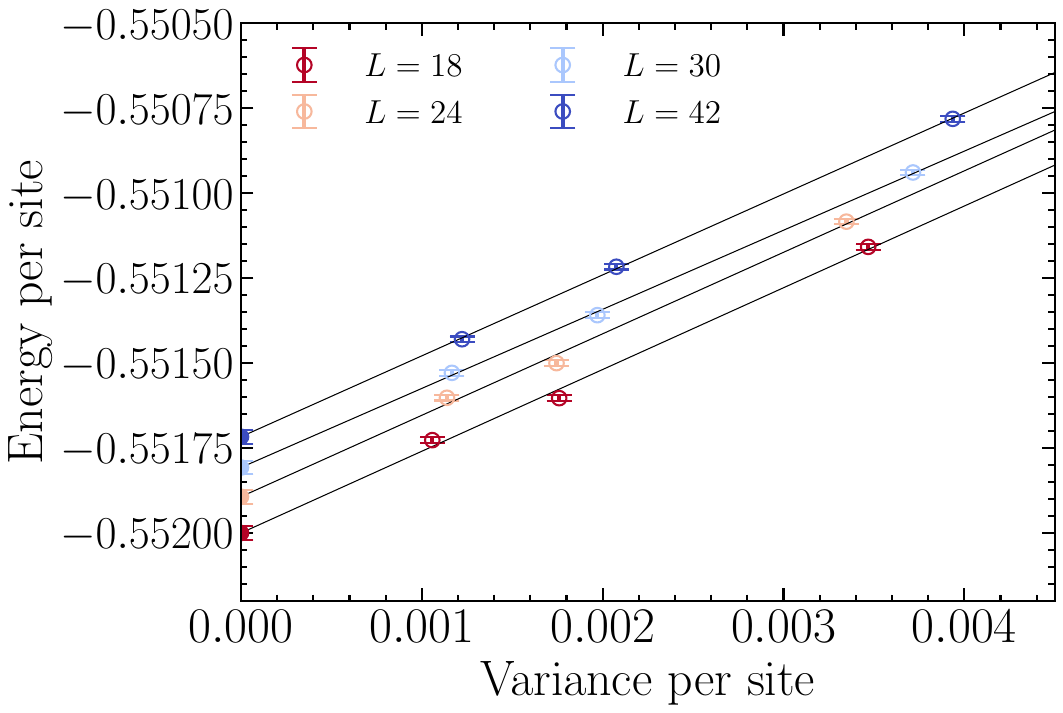}}
\caption{\label{fig:energy_variance_extrapolation}\textbf{Energy-variance extrapolations.} Variational energies as a function of the variance per site for the triangular-lattice Heisenberg model at system sizes $L=18,24,30$, and $42$. For each size, three points are shown corresponding to the unprojected ViT wave function, the state with translational symmetry restored, and the state with full point-group symmetry. Linear extrapolation to the zero-variance limit yields an estimate of the exact ground-state energy for each system size.}
    \end{center}
\end{figure}

\subsection{Architecture and Optimization Details}\label{sec:architecture}
The Vision Transformer architecture used in this work comprises $h=12$ attention heads, $n_l=8$ layers, and an embedding dimension of $d=72$, resulting in approximately $P \approx 4.5 \times10^{5}$ trainable parameters for different system sizes, from $L=18$ to $L=42$. The input spin configuration $\b\sigma$ is divided into $n$ non-overlapping patches of size $b\times b$~\cite{viteritti2025prb}, each embedded into $\mathbb{R}^{d}$ through a learnable linear transformation, producing the sequence $\b{x}_i$ with $i=1,\dots, n$, where $n=L^2/b^2$. In particular, we set $b=3$ for the triangular lattice, which is compatible with the Néel $120^{\circ}$ antiferromagnetic order and $b=4$ for the square lattice (see \cref{sec:square_lattice}). A detailed description of the role of the different hyperparameters of the architecture is provided in Ref.~\cite{viteritti2025prb}. We take the attention weights to be translationally invariant across patches and enforce all remaining lattice symmetries, including translations within patches, rotations, and reflections ($C_{6v}$ group for the triangular lattice and $C_{4v}$ for the square lattice) through a projector operator in the trivial representation sector~\cite{nomura2021helping}. The only modification to the standard architecture presented in Ref.~\cite{viteritti2025prb} is the inclusion of the Spatial Attention mechanism, where the variational parameter controlling the characteristic correlation range was initialized to $\gamma=3$ for all the heads and layers [see \cref{eq:distance_attn}]. Variational Monte Carlo (VMC) optimizations employing $M=2^{14}$ samples for the stochastic estimates were carried out using SR~\cite{sorella2005} with the linear algebra trick being in the regime $P \gg M$~\cite{chen2024empowering, rende2024stochastic} enhanced with SPRING~\cite{spring}. The optimization proceeded in three stages: $5000$ steps without symmetry summation, followed by $500$ steps including translational symmetries, and a final $100$ steps with both rotational and reflection symmetries restored. The initial learning rate was set to $\eta=5\times 10^{-3}$ and gradually decreased during training.

\begin{figure}[t]
    \begin{center}
\centerline{\includegraphics[width=0.9\columnwidth]{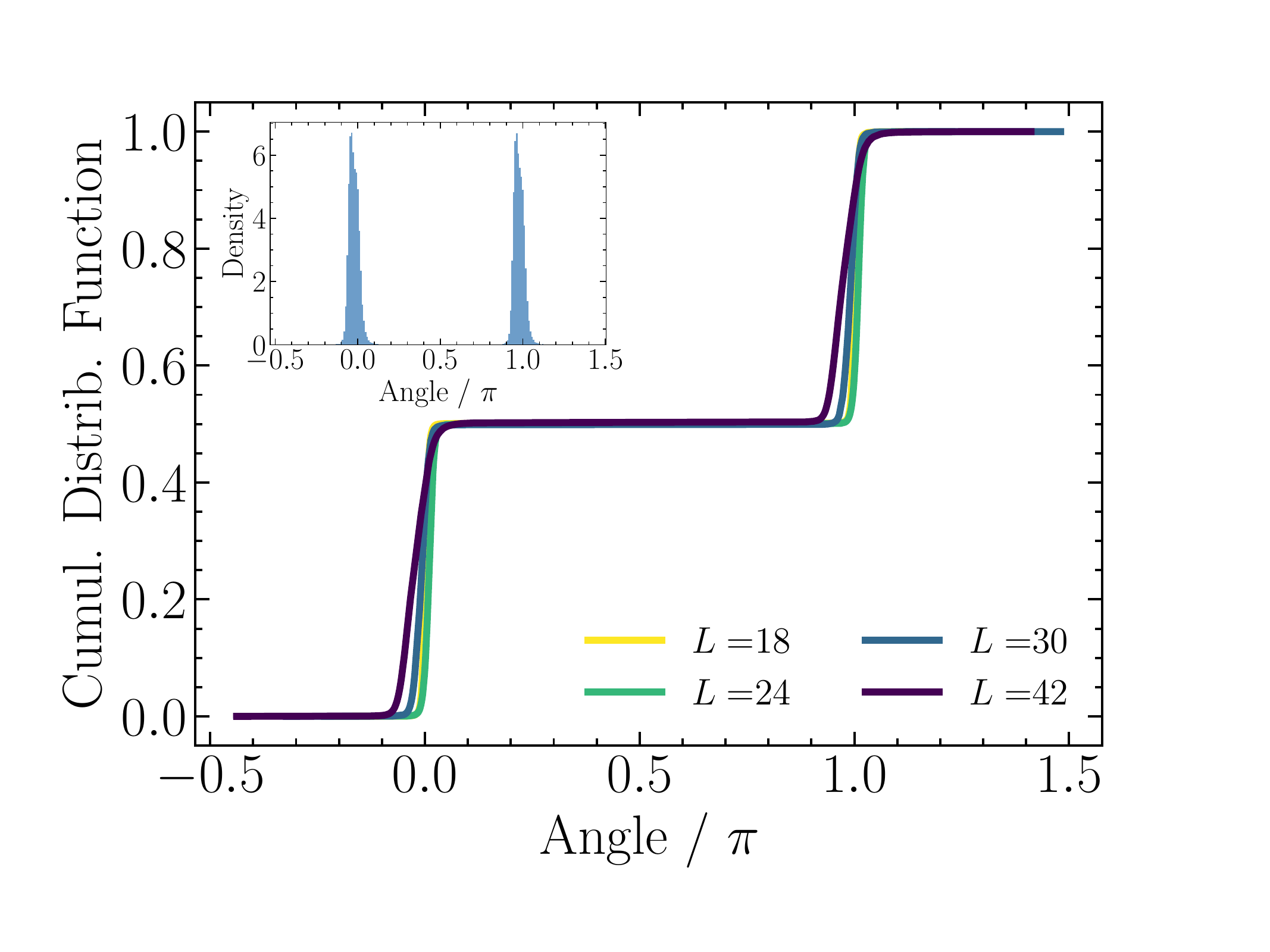}}
        \caption{\label{fig:cdf_sign}\textbf{Distribution of phases amplitude.} Cumulative distribution function of the phases of the neural-network amplitudes evaluated on samples drawn from the optimized variational state, shown for system sizes ${L=18,\,24,\,30}$, and $42$. \textbf{Inset:} Corresponding phases histogram for $L=42$.}
    \end{center}
\end{figure}

\subsection{Energy-Variance Extrapolations}\label{sec:en_var}
The eigenstates of a finite-size Hamiltonian can be classified according to the symmetry quantum numbers of the system. Although neural-network wave functions are universal approximators in the limit of sufficiently many parameters and can, in principle, represent any eigenstate of a given Hamiltonian, stochastic optimization does not automatically enforce these symmetries. As a result, the optimized variational state typically does not exactly satisfy the symmetry constraints.

Symmetry restoration can be performed \textit{a posteriori} through quantum-number projection~\cite{nomura2021helping, mizusaki2004}. This procedure not only enforces the exact symmetries of the Hamiltonian but also systematically improves the variational state, lowering the energy and reducing the variance. We can use this sequence of symmetry-projected states to perform a controlled variance extrapolation. Indeed, for a systematically improved sequence of states, it is easy to prove that
${\varepsilon_{\theta} \approx \varepsilon_{0} + \mathrm{const} \times \sigma_{\theta}^2}$,
where ${\varepsilon_{\theta} = {\langle \Psi_{\theta} | \hat{H} | \Psi_{\theta} \rangle}/L^2}$ and ${\sigma_{\theta}^2 = [{\langle \Psi_{\theta} | \hat{H}^2 | \Psi_{\theta} \rangle - \langle \Psi_{\theta} | \hat{H} | \Psi_{\theta} \rangle^2}]/{L^2}}$ are the energy and variance per site of the normalized state $|\Psi_{\theta}\rangle$~\cite{becca2013, becca2015}. The exact energy per site $\varepsilon_{0}$ can therefore be obtained by fitting $\varepsilon_{\theta}$ as a function of $\sigma_{\theta}^{2}$.

For the Heisenberg model on the triangular lattice, the sequence of improved states consists of three levels of symmetry restoration: the unprojected state, the state projected onto translational symmetry, and the state projected onto the full point-group symmetry ($C_{6v}$ group). As shown in \cref{fig:energy_variance_extrapolation}, the accuracy of the wave function improves systematically across this sequence, enabling a reliable zero-variance extrapolation for each system size from $L=18$ to $L=42$. We further emphasize that the variance per site remains remarkably stable as the system size increases, indicating that the variational Ansatz is size consistent (as discussed also in \cref{sec:energy_and_magn}).

\begin{figure}[t]
    \begin{center}
\centerline{\includegraphics[width=\columnwidth]{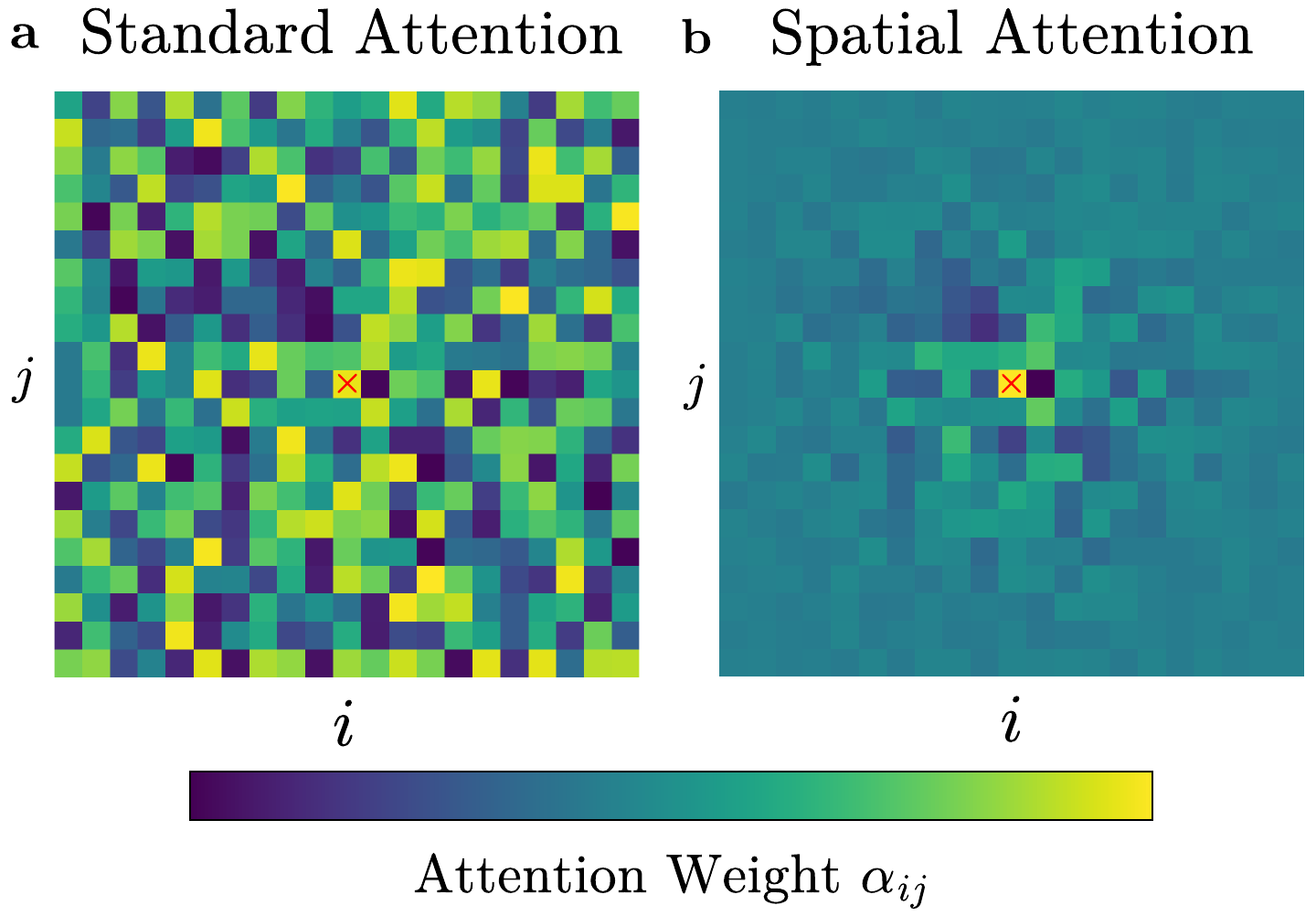}}
        \caption{\label{fig:attn_comparison} \textbf{Attention weights at initialization.} Attention weights $\alpha_{ij}$ \textit{at initialization} used to compute the updated representation of the central patch vector (marked by the red cross). \textbf{a.} Standard Factored Transformer attention~\cite{rende2025iop} assigns weights with no spatial structure, coupling the central patch uniformly to all others. \textbf{b.} Spatial attention introduces a learnable length scale that suppresses contributions from distant patches. Color intensity denotes the magnitude of $\alpha_{ij}$.}
    \end{center}
\end{figure}

\subsection{Distribution of Phases Amplitudes}\label{sec:phases_distrib}
Hamiltonians that respect time-reversal symmetry admit ground states that can be chosen to be real in the computational basis; equivalently, their many-body amplitudes carry phases restricted to $0$ or $\pi$, up to an overall global phase. Nonetheless, we work with complex-valued variational wavefunctions, as representing the sign structure in the complex plane substantially simplifies the optimization. However, we expect the optimized state to recover its real-valued character when converged~\cite{viteritti2022}.

In \cref{fig:cdf_sign} we report the cumulative distribution function of the phases of the neural-network amplitudes associated with configurations sampled from the optimized variational state. The comparison across system sizes, from $L=18$ up to $L=42$, shows that the phase distribution remains sharply concentrated around $0$ and $\pi$. While the width of these peaks increases mildly with system size, the overall distribution for $L=42$ remains strongly bimodal and extremely narrow, indicating that the optimized wavefunction is, to an excellent approximation, real-valued (see Inset of \cref{fig:cdf_sign}). \\

\subsection{Comparison of Standard and Spatial Factored Attention at initialization}
In \cref{fig:attn_comparison}, we compare the attention weights $\alpha_{ij}$ at initialization, which determine the contribution of each patch to the updated representation of the central patch (marked by a red cross), for both the Factored attention mechanism~\cite{rende2025iop} and the spatially weighted variant introduced in this work.
In the standard mechanism (see \cref{fig:attn_comparison}\textbf{a}), the attention weights $\alpha_{ij}$ show no spatial structure: all patches contribute with comparable strength. Such unstructured long-range parameters become increasingly challenging to optimize on large lattices. By contrast, with Spatial Attention (see \cref{fig:attn_comparison}\textbf{b}), the exponential kernel in \cref{eq:distance_attn} modulates the attention weights so that nearby patches receive systematically larger weights, while distant contributions are suppressed according to the learnable length scale $\gamma$. This embeds the spatial structure of the lattice directly at the architectural level, stabilizing the optimization for large system sizes.

\begin{acknowledgments}
We thank F. Becca, A. Georges, C. Batista, J. Hasik, S. Zhang, and A. Sandvik for useful discussions. We are grateful to A. Parola for drawing our attention to Ref.~\cite{Fisher1989}. We thank L. Reading-Ikkanda (Simons Foundation) for preparing \cref{fig:pictorial_representation}. LLV is supported
by SEFRI under Grant No. MB22.00051 (NEQS - Neural Quantum). This work was supported as part of the ``Swiss AI initiative'' by a grant from the Swiss National Supercomputing Centre (CSCS) under project ID a117 on Alps. RR and LLV acknowledge the CINECA award under the ISCRA initiative for the availability of high-performance computing resources and support. RR acknowledges support from the Flatiron Institute. The Flatiron Institute is a division of the Simons Foundation. The numerical experiments performed in this work required 25000 hours on GH200 GPUs. However, we point out that the preceding results obtained with other variational approaches, such as Gutzwiller Projected and RNN wave functions (see \cref{table:energies}), can be achieved with fewer than 1000 GPU hours.
\end{acknowledgments}

\bibliography{ref}

@article{roth2023,
  title = {High-accuracy variational Monte Carlo for frustrated magnets with deep neural networks},
  author = {Roth, Christopher and Szab\'o, Attila and MacDonald, Allan H.},
  journal = {Phys. Rev. B},
  volume = {108},
  issue = {5},
  pages = {054410},
  numpages = {12},
  year = {2023},
  month = {Aug},
  publisher = {American Physical Society},
  doi = {10.1103/PhysRevB.108.054410},
  url = {https://link.aps.org/doi/10.1103/PhysRevB.108.054410}
}

@article{lange2024,
doi = {10.1088/2058-9565/ad7168},
url = {https://dx.doi.org/10.1088/2058-9565/ad7168},
year = {2024},
month = {sep},
publisher = {IOP Publishing},
volume = {9},
number = {4},
pages = {040501},
author = {Lange, Hannah and Van de Walle, Anka and Abedinnia, Atiye and Bohrdt, Annabelle},
title = {From architectures to applications: a review of neural quantum states},
journal = {Quantum Science and Technology},
}

@article{clark2025prb,
  title = {Efficient optimization of neural network backflow for $ab initio$ quantum chemistry},
  author = {Liu, An-Jun and Clark, Bryan K.},
  journal = {Phys. Rev. B},
  volume = {112},
  issue = {15},
  pages = {155162},
  numpages = {17},
  year = {2025},
  month = {Oct},
  publisher = {American Physical Society},
  doi = {10.1103/thz7-lmdn},
  url = {https://link.aps.org/doi/10.1103/thz7-lmdn}
}

@article{liang2023,
   title={Deep learning representations for quantum many-body systems on heterogeneous hardware},
   volume={4},
   ISSN={2632-2153},
   url={http://dx.doi.org/10.1088/2632-2153/acc56a},
   DOI={10.1088/2632-2153/acc56a},
   number={1},
   journal={Machine Learning: Science and Technology},
   publisher={IOP Publishing},
   author={Liang, Xiao and Li, Mingfan and Xiao, Qian and Chen, Junshi and Yang, Chao and An, Hong and He, Lixin},
   year={2023},
   month=mar, pages={015035} }

@article{huang2024,
   title={On the magnetization of the 120° order of the spin-1/2 triangular lattice Heisenberg model: a DMRG revisited},
   volume={36},
   ISSN={1361-648X},
   url={http://dx.doi.org/10.1088/1361-648X/ad21a8},
   DOI={10.1088/1361-648x/ad21a8},
   number={18},
   journal={Journal of Physics: Condensed Matter},
   publisher={IOP Publishing},
   author={Huang, Jiale and Qian, Xiangjian and Qin, Mingpu},
   year={2024},
   month=feb, pages={185602} }

@article{rende2024stochastic,
   title={A simple linear algebra identity to optimize large-scale neural network quantum states},
   volume={7},
   ISSN={2399-3650},
   url={http://dx.doi.org/10.1038/s42005-024-01732-4},
   DOI={10.1038/s42005-024-01732-4},
   number={1},
   journal={Communications Physics},
   publisher={Springer Science and Business Media LLC},
   author={Rende, Riccardo and Viteritti, Luciano Loris and Bardone, Lorenzo and Becca, Federico and Goldt, Sebastian},
   year={2024},
   month=aug }

@article{gao2017,
   title={Efficient representation of quantum many-body states with deep neural networks},
   volume={8},
   ISSN={2041-1723},
   url={http://dx.doi.org/10.1038/s41467-017-00705-2},
   DOI={10.1038/s41467-017-00705-2},
   number={1},
   journal={Nature Communications},
   publisher={Springer Science and Business Media LLC},
   author={Gao, Xun and Duan, Lu-Ming},
   year={2017},
   month=sep }

@article{Sharir2019,
  title = {Quantum Entanglement in Deep Learning Architectures},
  author = {Levine, Yoav and Sharir, Or and Cohen, Nadav and Shashua, Amnon},
  journal = {Phys. Rev. Lett.},
  volume = {122},
  issue = {6},
  pages = {065301},
  numpages = {7},
  year = {2019},
  month = {Feb},
  publisher = {American Physical Society},
  doi = {10.1103/PhysRevLett.122.065301},
  url = {https://link.aps.org/doi/10.1103/PhysRevLett.122.065301}
}

@article{Sharir2022,
  title = {Neural tensor contractions and the expressive power of deep neural quantum states},
  author = {Sharir, Or and Shashua, Amnon and Carleo, Giuseppe},
  journal = {Phys. Rev. B},
  volume = {106},
  issue = {20},
  pages = {205136},
  numpages = {12},
  year = {2022},
  month = {Nov},
  publisher = {American Physical Society},
  doi = {10.1103/PhysRevB.106.205136},
  url = {https://link.aps.org/doi/10.1103/PhysRevB.106.205136}
}

@Article{viteritti2022,
  title={{Accuracy of restricted Boltzmann machines for the one-dimensional  $J_1-J_2$ Heisenberg model}},
  author={Viteritti, L.L. and Ferrari, F. and Becca, F.},
  journal={SciPost Phys.},
  volume={12},
  pages={166},
  year={2022},
  publisher={SciPost},
  doi={10.21468/SciPostPhys.12.5.166},
  url={https://scipost.org/10.21468/SciPostPhys.12.5.166},
}

@article{viteritti2023prl,
  title = {Transformer Variational Wave Functions for Frustrated Quantum Spin Systems},
  author = {Viteritti, Luciano Loris and Rende, Riccardo and Becca, Federico},
  journal = {Phys. Rev. Lett.},
  volume = {130},
  issue = {23},
  pages = {236401},
  numpages = {6},
  year = {2023},
  month = {Jun},
  publisher = {American Physical Society},
  doi = {10.1103/PhysRevLett.130.236401},
  url = {https://link.aps.org/doi/10.1103/PhysRevLett.130.236401}
}

@article{szabo2020,
  title = {Neural network wave functions and the sign problem},
  author = {Szab\'o, Attila and Castelnovo, Claudio},
  journal = {Phys. Rev. Res.},
  volume = {2},
  issue = {3},
  pages = {033075},
  numpages = {12},
  year = {2020},
  month = {Jul},
  publisher = {American Physical Society},
  doi = {10.1103/PhysRevResearch.2.033075},
  url = {https://link.aps.org/doi/10.1103/PhysRevResearch.2.033075}
}

@article{viteritti2025prb,
  title = {Transformer wave function for two dimensional frustrated magnets: Emergence of a spin-liquid phase in the Shastry-Sutherland model},
  author = {Viteritti, Luciano Loris and Rende, Riccardo and Parola, Alberto and Goldt, Sebastian and Becca, Federico},
  journal = {Phys. Rev. B},
  volume = {111},
  issue = {13},
  pages = {134411},
  numpages = {15},
  year = {2025},
  month = {Apr},
  publisher = {American Physical Society},
  doi = {10.1103/PhysRevB.111.134411},
  url = {https://link.aps.org/doi/10.1103/PhysRevB.111.134411}
}

@article{rende2024prr,
  title = {Mapping of attention mechanisms to a generalized Potts model},
  author = {Rende, Riccardo and Gerace, Federica and Laio, Alessandro and Goldt, Sebastian},
  journal = {Phys. Rev. Res.},
  volume = {6},
  issue = {2},
  pages = {023057},
  numpages = {10},
  year = {2024},
  month = {Apr},
  publisher = {American Physical Society},
  doi = {10.1103/PhysRevResearch.6.023057},
  url = {https://link.aps.org/doi/10.1103/PhysRevResearch.6.023057}
}

@inproceedings{rende2024neurips,
 author = {Rende, Riccardo and Gerace, Federica and Laio, Alessandro and Goldt, Sebastian},
 booktitle = {Advances in Neural Information Processing Systems},
 editor = {A. Globerson and L. Mackey and D. Belgrave and A. Fan and U. Paquet and J. Tomczak and C. Zhang},
 pages = {96207--96228},
 publisher = {Curran Associates, Inc.},
 title = {A distributional simplicity bias in the learning dynamics of transformers},
 url = {https://proceedings.neurips.cc/paper_files/paper/2024/file/ae6c81a39079ddeb88b034b6ef18c7fe-Paper-Conference.pdf},
 volume = {37},
 year = {2024}
}

@article{rende2025iop,
doi = {10.1088/2632-2153/ada1a0},
url = {https://doi.org/10.1088/2632-2153/ada1a0},
year = {2025},
month = {jan},
publisher = {IOP Publishing},
volume = {6},
number = {1},
pages = {010501},
author = {Rende, Riccardo and Loris Viteritti, Luciano},
title = {Are queries and keys always relevant? A case study on transformer wave functions},
journal = {Machine Learning: Science and Technology}
}

@article{sorellaprb2006,
  title = {Two spin liquid phases in the spatially anisotropic triangular Heisenberg model},
  author = {Yunoki, Seiji and Sorella, Sandro},
  journal = {Phys. Rev. B},
  volume = {74},
  issue = {1},
  pages = {014408},
  numpages = {31},
  year = {2006},
  month = {Jul},
  publisher = {American Physical Society},
  doi = {10.1103/PhysRevB.74.014408},
  url = {https://link.aps.org/doi/10.1103/PhysRevB.74.014408}
}

@article{beccaprb2009,
  title = {Spin-$\frac{1}{2}$ Heisenberg model on the anisotropic triangular lattice: From magnetism to a one-dimensional spin liquid},
  author = {Heidarian, Dariush and Sorella, Sandro and Becca, Federico},
  journal = {Phys. Rev. B},
  volume = {80},
  issue = {1},
  pages = {012404},
  numpages = {4},
  year = {2009},
  month = {Jul},
  publisher = {American Physical Society},
  doi = {10.1103/PhysRevB.80.012404},
  url = {https://link.aps.org/doi/10.1103/PhysRevB.80.012404}
}

@article{beccaprb2016,
  title = {Variational wave functions for the $S=\frac{1}{2}$ Heisenberg model on the anisotropic triangular lattice: Spin liquids and spiral orders},
  author = {Ghorbani, Elaheh and Tocchio, Luca F. and Becca, Federico},
  journal = {Phys. Rev. B},
  volume = {93},
  issue = {8},
  pages = {085111},
  numpages = {10},
  year = {2016},
  month = {Feb},
  publisher = {American Physical Society},
  doi = {10.1103/PhysRevB.93.085111},
  url = {https://link.aps.org/doi/10.1103/PhysRevB.93.085111}
}

@misc{moss2025rnn,
      title={Leveraging recurrence in neural network wavefunctions for large-scale simulations of Heisenberg antiferromagnets on the triangular lattice}, 
      author={M. Schuyler Moss and Roeland Wiersema and Mohamed Hibat-Allah and Juan Carrasquilla and Roger G. Melko},
      year={2025},
      eprint={2505.20406},
      archivePrefix={arXiv},
      primaryClass={cond-mat.str-el},
      url={https://arxiv.org/abs/2505.20406}, 
}

@article{mezzacapo2010iop,
doi = {10.1088/1367-2630/12/10/103039},
url = {https://doi.org/10.1088/1367-2630/12/10/103039},
year = {2010},
month = {oct},
publisher = {},
volume = {12},
number = {10},
pages = {103039},
author = {Mezzacapo, F and Cirac, J I},
title = {Ground-state properties of the spin- antiferromagnetic Heisenberg model on the triangular lattice: a variational study based on entangled-plaquette states},
journal = {New Journal of Physics},
}

@misc{verstraete2004,
      title={Renormalization algorithms for Quantum-Many Body Systems in two and higher dimensions}, 
      author={F. Verstraete and J. I. Cirac},
      year={2004},
      eprint={cond-mat/0407066},
      archivePrefix={arXiv},
      primaryClass={cond-mat.str-el},
      url={https://arxiv.org/abs/cond-mat/0407066}, 
}

@article{Jordan2008,
   title={Classical Simulation of Infinite-Size Quantum Lattice Systems in Two Spatial Dimensions},
   volume={101},
   ISSN={1079-7114},
   url={http://dx.doi.org/10.1103/PhysRevLett.101.250602},
   DOI={10.1103/physrevlett.101.250602},
   number={25},
   journal={Physical Review Letters},
   publisher={American Physical Society (APS)},
   author={Jordan, J. and Orús, R. and Vidal, G. and Verstraete, F. and Cirac, J. I.},
   year={2008},
   month=dec }

@article{Corboz2016,
   title={Variational optimization with infinite projected entangled-pair states},
   volume={94},
   ISSN={2469-9969},
   url={http://dx.doi.org/10.1103/PhysRevB.94.035133},
   DOI={10.1103/physrevb.94.035133},
   number={3},
   journal={Physical Review B},
   publisher={American Physical Society (APS)},
   author={Corboz, Philippe},
   year={2016},
   month=jul }

@article{Corboz2012,
  title = {Spin-Orbital Quantum Liquid on the Honeycomb Lattice},
  author = {Corboz, Philippe and Lajk\'o, Mikl\'os and L\"auchli, Andreas M. and Penc, Karlo and Mila, Fr\'ed\'eric},
  journal = {Phys. Rev. X},
  volume = {2},
  issue = {4},
  pages = {041013},
  numpages = {11},
  year = {2012},
  month = {Nov},
  publisher = {American Physical Society},
  doi = {10.1103/PhysRevX.2.041013},
  url = {https://link.aps.org/doi/10.1103/PhysRevX.2.041013}
}

@article{Vlaar2021,
  title = {Simulation of three-dimensional quantum systems with projected entangled-pair states},
  author = {Vlaar, Patrick C. G. and Corboz, Philippe},
  journal = {Phys. Rev. B},
  volume = {103},
  issue = {20},
  pages = {205137},
  numpages = {11},
  year = {2021},
  month = {May},
  publisher = {American Physical Society},
  doi = {10.1103/PhysRevB.103.205137},
  url = {https://link.aps.org/doi/10.1103/PhysRevB.103.205137}
}

@article{Scheie2024,
  title = {Nonlinear magnons and exchange Hamiltonians of the delafossite proximate quantum spin liquid candidates ${\text{KYbSe}}_{2}$ and ${\text{NaYbSe}}_{2}$},
  author = {Scheie, A. O. and Kamiya, Y. and Zhang, Hao and Lee, Sangyun and Woods, A. J. and Ajeesh, M. O. and Gonzalez, M. G. and Bernu, B. and Villanova, J. W. and Xing, J. and Huang, Q. and Zhang, Qingming and Ma, Jie and Choi, Eun Sang and Pajerowski, D. M. and Zhou, Haidong and Sefat, A. S. and Okamoto, S. and Berlijn, T. and Messio, L. and Movshovich, R. and Batista, C. D. and Tennant, D. A.},
  journal = {Phys. Rev. B},
  volume = {109},
  issue = {1},
  pages = {014425},
  numpages = {12},
  year = {2024},
  month = {Jan},
  publisher = {American Physical Society},
  doi = {10.1103/PhysRevB.109.014425},
  url = {https://link.aps.org/doi/10.1103/PhysRevB.109.014425}
}

@article{Susuki2013,
  title = {Magnetization Process and Collective Excitations in the $S\mathbf{=}1/2$ Triangular-Lattice Heisenberg Antiferromagnet ${\mathrm{Ba}}_{3}{\mathrm{CoSb}}_{2}{\mathbf{O}}_{9}$},
  author = {Susuki, Takuya and Kurita, Nobuyuki and Tanaka, Takuya and Nojiri, Hiroyuki and Matsuo, Akira and Kindo, Koichi and Tanaka, Hidekazu},
  journal = {Phys. Rev. Lett.},
  volume = {110},
  issue = {26},
  pages = {267201},
  numpages = {5},
  year = {2013},
  month = {Jun},
  publisher = {American Physical Society},
  doi = {10.1103/PhysRevLett.110.267201},
  url = {https://link.aps.org/doi/10.1103/PhysRevLett.110.267201}
}

@misc{solinas2026,
      title={Neural Quantum States in Mixed Precision}, 
      author={Massimo Solinas and Agnes Valenti and Nawaf Bou-Rabee and Roeland Wiersema},
      year={2026},
      eprint={2601.20782},
      archivePrefix={arXiv},
      primaryClass={quant-ph},
      url={https://arxiv.org/abs/2601.20782}, 
}

@misc{shokry2025,
      title={When Less is More: Approximating the Quantum Geometric Tensor with Block Structures}, 
      author={Ahmedeo Shokry and Alessandro Santini and Filippo Vicentini},
      year={2025},
      eprint={2510.08430},
      archivePrefix={arXiv},
      primaryClass={quant-ph},
      url={https://arxiv.org/abs/2510.08430}, 
}

@misc{misery2025,
      title={Looking elsewhere: improving variational Monte Carlo gradients by importance sampling}, 
      author={Antoine Misery and Luca Gravina and Alessandro Santini and Filippo Vicentini},
      year={2025},
      eprint={2507.05352},
      archivePrefix={arXiv},
      primaryClass={quant-ph},
      url={https://arxiv.org/abs/2507.05352}, 
}

@article{LiebRobinson1972,
  author       = {Lieb, E. H. and Robinson, D. W.},
  title        = {The finite group velocity of quantum spin systems},
  journal      = {Commun. Math. Phys.},
  volume       = {28},
  pages        = {251--257},
  year         = {1972},
  doi          = {10.1007/BF01645779},
  url          = {https://doi.org/10.1007/BF01645779}
}

@article{HastingsKoma2006,
  author       = {Hastings, M. B. and Koma, T.},
  title        = {Spectral gap and exponential decay of correlations},
  journal      = {Commun. Math. Phys.},
  volume       = {265},
  pages        = {781--804},
  year         = {2006},
  doi          = {10.1007/s00220-006-0030-4},
  url          = {https://doi.org/10.1007/s00220-006-0030-4}
}

@article{sandvik1999,
  title = {Stochastic series expansion method with operator-loop update},
  author = {Sandvik, Anders W.},
  journal = {Phys. Rev. B},
  volume = {59},
  issue = {22},
  pages = {R14157--R14160},
  numpages = {0},
  year = {1999},
  month = {Jun},
  publisher = {American Physical Society},
  doi = {10.1103/PhysRevB.59.R14157},
  url = {https://link.aps.org/doi/10.1103/PhysRevB.59.R14157}
}

@article{sandvik1997,
  title = {Finite-size scaling of the ground-state parameters of the two-dimensional Heisenberg model},
  author = {Sandvik, Anders W.},
  journal = {Phys. Rev. B},
  volume = {56},
  issue = {18},
  pages = {11678--11690},
  numpages = {0},
  year = {1997},
  month = {Nov},
  publisher = {American Physical Society},
  doi = {10.1103/PhysRevB.56.11678},
  url = {https://link.aps.org/doi/10.1103/PhysRevB.56.11678}
}

@article{igarashi1992,
  title = {1/S expansion for thermodynamic quantities in a two-dimensional Heisenberg antiferromagnet at zero temperature},
  author = {Igarashi, Jun-ichi},
  journal = {Phys. Rev. B},
  volume = {46},
  issue = {17},
  pages = {10763--10771},
  numpages = {0},
  year = {1992},
  month = {Nov},
  publisher = {American Physical Society},
  doi = {10.1103/PhysRevB.46.10763},
  url = {https://link.aps.org/doi/10.1103/PhysRevB.46.10763}
}

@article{chubukov1999,
author = {Chubukov, Andrey and Sachdev, Subir and Todadri, Senthil},
year = {1999},
month = {01},
pages = {8891},
title = {Large-S expansion for quantum antiferromagnets on a triangular lattice},
volume = {6},
journal = {Journal of Physics: Condensed Matter},
doi = {10.1088/0953-8984/6/42/019}
}

@article{Chubukov1994,
  title = {Universal magnetic properties of frustrated quantum antiferromagnets in two dimensions},
  author = {Chubukov, Andrey V. and Senthil, T. and Sachdev, Subir},
  journal = {Phys. Rev. Lett.},
  volume = {72},
  issue = {13},
  pages = {2089--2092},
  numpages = {0},
  year = {1994},
  month = {Mar},
  publisher = {American Physical Society},
  doi = {10.1103/PhysRevLett.72.2089},
  url = {https://link.aps.org/doi/10.1103/PhysRevLett.72.2089}
}

@article{Chubukov1994a,
   title={Quantum phase transitions in frustrated quantum antiferromagnets},
   volume={426},
   ISSN={0550-3213},
   url={http://dx.doi.org/10.1016/0550-3213(94)90023-X},
   DOI={10.1016/0550-3213(94)90023-x},
   number={3},
   journal={Nuclear Physics B},
   publisher={Elsevier BV},
   author={Chubukov, Andrey V. and Sachdev, Subir and Senthil, T.},
   year={1994},
   month=sep, pages={601–643} }

@article{chernyshev2009,
  title = {Spin waves in a triangular lattice antiferromagnet: Decays, spectrum renormalization, and singularities},
  author = {Chernyshev, A. L. and Zhitomirsky, M. E.},
  journal = {Phys. Rev. B},
  volume = {79},
  issue = {14},
  pages = {144416},
  numpages = {28},
  year = {2009},
  month = {Apr},
  publisher = {American Physical Society},
  doi = {10.1103/PhysRevB.79.144416},
  url = {https://link.aps.org/doi/10.1103/PhysRevB.79.144416}
}

@book{becca2017,
  title = {Quantum Monte Carlo Approaches for Correlated Systems},
  author = {Becca, F. and Sorella, S.},
  publisher = {Cambridge University Press},
  year = {2017},
  doi = {10.1017/9781316417041},
  url = {https://www.cambridge.org/core/books/quantum-monte-carlo-approaches-for-correlated-systems/EB88C86BD9553A0738BDAE400D0B2900}
}

@article{rende2025foundations,
   title={Foundation neural-networks quantum states as a unified Ansatz for multiple hamiltonians},
   volume={16},
   ISSN={2041-1723},
   url={http://dx.doi.org/10.1038/s41467-025-62098-x},
   DOI={10.1038/s41467-025-62098-x},
   number={1},
   journal={Nature Communications},
   publisher={Springer Science and Business Media LLC},
   author={Rende, Riccardo and Viteritti, Luciano Loris and Becca, Federico and Scardicchio, Antonello and Laio, Alessandro and Carleo, Giuseppe},
   year={2025},
   month=aug }

@misc{viteritti2025spinglass,
      title={Quantum Spin Glass in the Two-Dimensional Disordered Heisenberg Model via Foundation Neural-Network Quantum States}, 
      author={Luciano Loris Viteritti and Riccardo Rende and Giacomo Bracci Testasecca and Jacopo Niedda and Roderich Moessner and Giuseppe Carleo and Antonello Scardicchio},
      year={2025},
      eprint={2507.05073},
      archivePrefix={arXiv},
      primaryClass={cond-mat.dis-nn},
      url={https://arxiv.org/abs/2507.05073}, 
}

@article{Nutakki2025,
   title={Design principles of deep translationally symmetric neural quantum states for frustrated magnets},
   volume={7},
   ISSN={2643-1564},
   url={http://dx.doi.org/10.1103/ybgv-35jm},
   DOI={10.1103/ybgv-35jm},
   number={4},
   journal={Physical Review Research},
   publisher={American Physical Society (APS)},
   author={Nutakki, Rajah P. and Shokry, Ahmedeo and Vicentini, Filippo},
   year={2025},
   month=oct }

@misc{chen2025,
      title={Convolutional transformer wave functions}, 
      author={Ao Chen and Vighnesh Dattatraya Naik and Markus Heyl},
      year={2025},
      eprint={2503.10462},
      archivePrefix={arXiv},
      primaryClass={cond-mat.dis-nn},
      url={https://arxiv.org/abs/2503.10462}, 
}

@misc{wu2021,
      title={DA-Transformer: Distance-aware Transformer}, 
      author={Chuhan Wu and Fangzhao Wu and Yongfeng Huang},
      year={2021},
      eprint={2010.06925},
      archivePrefix={arXiv},
      primaryClass={cs.CL},
      url={https://arxiv.org/abs/2010.06925}, 
}

@misc{zhang2025,
      title={Interpretable Artificial Intelligence (AI) Analysis of Strongly Correlated Electrons}, 
      author={Changkai Zhang and Jan von Delft},
      year={2025},
      eprint={2510.26864},
      archivePrefix={arXiv},
      primaryClass={cond-mat.str-el},
      url={https://arxiv.org/abs/2510.26864}, 
}

@misc{sun2022,
      title={Locality Matters: A Locality-Biased Linear Attention for Automatic Speech Recognition}, 
      author={Jingyu Sun and Guiping Zhong and Dinghao Zhou and Baoxiang Li and Yiran Zhong},
      year={2022},
      eprint={2203.15609},
      archivePrefix={arXiv},
      primaryClass={cs.SD},
      url={https://arxiv.org/abs/2203.15609}, 
}

@article{Anderson1987,
  author       = {Anderson, P. W.},
  title        = {The resonating valence bond state in La$_2$CuO$_4$ and superconductivity},
  journal      = {Science},
  volume       = {235},
  pages        = {1196--1198},
  year         = {1987},
  doi          = {10.1126/science.235.4793.1196},
  url          = {https://doi.org/10.1126/science.235.4793.1196}
}

@book{SachdevBook,
  author       = {Sachdev, Subir},
  title        = {Quantum Phase Transitions},
  edition      = {2},
  publisher    = {Cambridge University Press},
  year         = {2011},
  doi          = {10.1017/CBO9780511973765},
  url          = {https://doi.org/10.1017/CBO9780511973765}
}

@article{TroyerWiese2005,
  author       = {Troyer, M. and Wiese, U.-J.},
  title        = {Computational Complexity and Fundamental Limitations to Fermionic Quantum Monte Carlo Simulations},
  journal      = {Phys. Rev. Lett.},
  volume       = {94},
  pages        = {170201},
  year         = {2005},
  doi          = {10.1103/PhysRevLett.94.170201},
  url          = {https://doi.org/10.1103/PhysRevLett.94.170201}
}

@article{Balents2010,
  author       = {Balents, L.},
  title        = {Spin liquids in frustrated magnets},
  journal      = {Nature},
  volume       = {464},
  pages        = {199--208},
  year         = {2010},
  doi          = {10.1038/nature08917},
  url          = {https://doi.org/10.1038/nature08917}
}

@article{EisertCramerPlenio2010_Entanglement,
  author       = {Eisert, J. and Cramer, M. and Plenio, M. B.},
  title        = {Area laws for the entanglement entropy},
  journal      = {Rev. Mod. Phys.},
  volume       = {82},
  number       = {1},
  pages        = {277--306},
  year         = {2010},
  doi          = {10.1103/RevModPhys.82.277},
  url          = {https://doi.org/10.1103/RevModPhys.82.277}
}

@article{rende2024fine,
   title={Fine-tuning neural network quantum states},
   volume={6},
   ISSN={2643-1564},
   url={http://dx.doi.org/10.1103/PhysRevResearch.6.043280},
   DOI={10.1103/physrevresearch.6.043280},
   number={4},
   journal={Physical Review Research},
   publisher={American Physical Society (APS)},
   author={Rende, Riccardo and Goldt, Sebastian and Becca, Federico and Viteritti, Luciano Loris},
   year={2024},
   month=dec }

@article{czischek2023,
  title={Variational Monte Carlo with Large Patched Transformers}, 
  author={Kyle Sprague and Stefanie Czischek},
  journal={arXiv preprint arXiv:2306.03921},
  year={2023},
  eprint={2306.03921},
  archivePrefix={arXiv},
  primaryClass={quant-ph}
}

@article{Schollwock2011,
  author       = {Schollw{\"o}ck, U.},
  title        = {The density-matrix renormalization group in the age of matrix product states},
  journal      = {Ann. Phys.},
  volume       = {326},
  pages        = {96--192},
  year         = {2011},
  doi          = {10.1016/j.aop.2010.09.012},
  url          = {https://doi.org/10.1016/j.aop.2010.09.012}
}

@article{nomura2021,
  title = {Dirac-Type Nodal Spin Liquid Revealed by Refined Quantum Many-Body Solver Using Neural-Network Wave Function, Correlation Ratio, and Level Spectroscopy},
  author = {Nomura, Yusuke and Imada, Masatoshi},
  journal = {Phys. Rev. X},
  volume = {11},
  issue = {3},
  pages = {031034},
  numpages = {19},
  year = {2021},
  month = {Aug},
  publisher = {American Physical Society},
  doi = {10.1103/PhysRevX.11.031034},
  url = {https://link.aps.org/doi/10.1103/PhysRevX.11.031034}
}

@article{nomura2021helping,
   title={Helping restricted Boltzmann machines with quantum-state representation by restoring symmetry},
   volume={33},
   ISSN={1361-648X},
   url={http://dx.doi.org/10.1088/1361-648X/abe268},
   DOI={10.1088/1361-648x/abe268},
   number={17},
   journal={Journal of Physics: Condensed Matter},
   publisher={IOP Publishing},
   author={Nomura, Yusuke},
   year={2021},
   month=apr, pages={174003} }

@article{mizusaki2004,
  title = {Quantum-number projection in the path-integral renormalization group method},
  author = {Mizusaki, Takahiro and Imada, Masatoshi},
  journal = {Phys. Rev. B},
  volume = {69},
  issue = {12},
  pages = {125110},
  numpages = {10},
  year = {2004},
  month = {Mar},
  publisher = {American Physical Society},
  doi = {10.1103/PhysRevB.69.125110},
  url = {https://link.aps.org/doi/10.1103/PhysRevB.69.125110}
}

@article{becca2015,
   title={Lanczos steps to improve variational wave functions},
   volume={640},
   ISSN={1742-6596},
   url={http://dx.doi.org/10.1088/1742-6596/640/1/012039},
   DOI={10.1088/1742-6596/640/1/012039},
   journal={Journal of Physics: Conference Series},
   publisher={IOP Publishing},
   author={Becca, Federico and Hu, Wen-Jun and Iqbal, Yasir and Parola, Alberto and Poilblanc, Didier and Sorella, Sandro},
   year={2015},
   month=sep, pages={012039} }

@article{becca2013,
  title = {Direct evidence for a gapless ${Z}_{2}$ spin liquid by frustrating N\'eel antiferromagnetism},
  author = {Hu, Wen-Jun and Becca, Federico and Parola, Alberto and Sorella, Sandro},
  journal = {Phys. Rev. B},
  volume = {88},
  issue = {6},
  pages = {060402},
  numpages = {5},
  year = {2013},
  month = {Aug},
  publisher = {American Physical Society},
  doi = {10.1103/PhysRevB.88.060402},
  url = {https://link.aps.org/doi/10.1103/PhysRevB.88.060402}
}

@misc{vaswani2023,
      title={Attention Is All You Need}, 
      author={Ashish Vaswani and Noam Shazeer and Niki Parmar and Jakob Uszkoreit and Llion Jones and Aidan N. Gomez and Lukasz Kaiser and Illia Polosukhin},
      year={2023},
      eprint={1706.03762},
      archivePrefix={arXiv},
      primaryClass={cs.CL},
      url={https://arxiv.org/abs/1706.03762}, 
}

@article{NeubergerZiman1989,
  author       = {Neuberger, Herbert and Ziman, Timothy},
  title        = {Finite-size effects in Heisenberg antiferromagnets},
  journal      = {Phys. Rev. B},
  volume       = {39},
  pages        = {2608--2618},
  year         = {1989},
  doi          = {10.1103/PhysRevB.39.2608},
  url          = {https://doi.org/10.1103/PhysRevB.39.2608}
}

@inproceedings{Sandvik2010,
  author       = {Sandvik, Anders W.},
  title        = {Computational Studies of Quantum Spin Systems},
  booktitle    = {AIP Conference Proceedings},
  volume       = {1297},
  pages        = {135--338},
  year         = {2010},
  doi          = {10.1063/1.3518900},
  url          = {https://doi.org/10.1063/1.3518900},
  publisher    = {AIP Publishing}
}

@article{chernyshev2007,
  title = {Ne\'el Order in Square and Triangular Lattice Heisenberg Models},
  author = {White, Steven R. and Chernyshev, A. L.},
  journal = {Phys. Rev. Lett.},
  volume = {99},
  issue = {12},
  pages = {127004},
  numpages = {4},
  year = {2007},
  month = {Sep},
  publisher = {American Physical Society},
  doi = {10.1103/PhysRevLett.99.127004},
  url = {https://link.aps.org/doi/10.1103/PhysRevLett.99.127004}
}

@article{batista2016,
  title = {Static and Dynamical Properties of the Spin-$1/2$ Equilateral Triangular-Lattice Antiferromagnet ${\mathrm{Ba}}_{3}{\mathrm{CoSb}}_{2}{\mathrm{O}}_{9}$},
  author = {Ma, J. and Kamiya, Y. and Hong, Tao and Cao, H. B. and Ehlers, G. and Tian, W. and Batista, C. D. and Dun, Z. L. and Zhou, H. D. and Matsuda, M.},
  journal = {Phys. Rev. Lett.},
  volume = {116},
  issue = {8},
  pages = {087201},
  numpages = {5},
  year = {2016},
  month = {Feb},
  publisher = {American Physical Society},
  doi = {10.1103/PhysRevLett.116.087201},
  url = {https://link.aps.org/doi/10.1103/PhysRevLett.116.087201}
}

@article{xie2023,
  title = {Complete field-induced spectral response of the spin-1/2 triangular-lattice antiferromagnet CsYbSe$_2$},
  author = {Xie, Tao and Eberharter, A. A. and Xing, Jie and Nishimoto, S. and Brando, M. and Khanenko, P. and Sichelschmidt, J. and Turrini, A. A. and Mazzone, D. G. and Naumov, P. G. and Sanjeewa, L. D. and Harrison, N. and Sefat, Athena S. and Normand, B. and Läuchli, A. M. and Podlesnyak, A. and Nikitin, S. E.},
  journal = {npj Quantum Materials},
  volume = {8},
  year = {2023},
  pages = {48},
  doi = {10.1038/s41535-023-00580-9}
}

@article{scheie2023,
  title = {Proximate spin liquid and fractionalization in the triangular antiferromagnet KYbSe$_2$},
  author = {Scheie, A. O. and Ghioldi, E. A. and Xing, J. and Paddison, J. A. M. and Sherman, N. E. and Dupont, M. and Sanjeewa, L. D. and Lee, Sangyun and Woods, A. J. and Abernathy, D. and Pajerowski, D. M. and Williams, T. J. and Zhang, Shang-Shun and Manuel, L. O. and Trumper, A. E. and Pemmaraju, C. D. and Sefat, A. S. and Parker, D. S. and Devereaux, T. P. and Movshovich, R. and Moore, J. E. and Batista, C. D. and Tennant, D. A.},
  journal = {Nature Physics},
  volume = {20},
  pages = {74--81},
  year = {2024},
  doi = {10.1038/s41567-023-02259-1}
}

@article{sorella2005,
  title = {Wave function optimization in the variational Monte Carlo method},
  author = {Sorella, Sandro},
  journal = {Phys. Rev. B},
  volume = {71},
  issue = {24},
  pages = {241103},
  numpages = {4},
  year = {2005},
  month = {Jun},
  publisher = {American Physical Society},
  doi = {10.1103/PhysRevB.71.241103},
  url = {https://link.aps.org/doi/10.1103/PhysRevB.71.241103}
}

@article{torlai2020,
  title = {Wave-function positivization via automatic differentiation},
  author = {Torlai, Giacomo and Carrasquilla, Juan and Fishman, Matthew T. and Melko, Roger G. and Fisher, Matthew P. A.},
  journal = {Phys. Rev. Res.},
  volume = {2},
  issue = {3},
  pages = {032060},
  numpages = {5},
  year = {2020},
  month = {Sep},
  publisher = {American Physical Society},
  doi = {10.1103/PhysRevResearch.2.032060},
  url = {https://link.aps.org/doi/10.1103/PhysRevResearch.2.032060}
}

@misc{naumann2025,
      title={Variational optimization of projected entangled-pair states on the triangular lattice}, 
      author={Jan Naumann and Jens Eisert and Philipp Schmoll},
      year={2025},
      eprint={2510.04907},
      archivePrefix={arXiv},
      primaryClass={cond-mat.str-el},
      url={https://arxiv.org/abs/2510.04907}, 
}

@article{chi2022,
  title = {Spin Excitation Spectra of Anisotropic Spin-$1/2$ Triangular Lattice Heisenberg Antiferromagnets},
  author = {Chi, Runze and Liu, Yang and Wan, Yuan and Liao, Hai-Jun and Xiang, T.},
  journal = {Phys. Rev. Lett.},
  volume = {129},
  issue = {22},
  pages = {227201},
  numpages = {6},
  year = {2022},
  month = {Nov},
  publisher = {American Physical Society},
  doi = {10.1103/PhysRevLett.129.227201},
  url = {https://link.aps.org/doi/10.1103/PhysRevLett.129.227201}
}

@article{trumper2000,
  title = {Finite-size spin-wave theory of the triangular Heisenberg model},
  author = {Trumper, Adolfo E. and Capriotti, Luca and Sorella, Sandro},
  journal = {Phys. Rev. B},
  volume = {61},
  issue = {17},
  pages = {11529--11532},
  numpages = {0},
  year = {2000},
  month = {May},
  publisher = {American Physical Society},
  doi = {10.1103/PhysRevB.61.11529},
  url = {https://link.aps.org/doi/10.1103/PhysRevB.61.11529}
}

@misc{gu2025,
      title={Solving the Hubbard model with Neural Quantum States}, 
      author={Yuntian Gu and Wenrui Li and Heng Lin and Bo Zhan and Ruichen Li and Yifei Huang and Di He and Yantao Wu and Tao Xiang and Mingpu Qin and Liwei Wang and Dingshun Lv},
      year={2025},
      eprint={2507.02644},
      archivePrefix={arXiv},
      primaryClass={cond-mat.str-el},
      url={https://arxiv.org/abs/2507.02644}, 
}

@article{
vscore,
author = {Dian Wu  and Riccardo Rossi  and Filippo Vicentini  and Nikita Astrakhantsev  and Federico Becca  and Xiaodong Cao  and Juan Carrasquilla  and Francesco Ferrari  and Antoine Georges  and Mohamed Hibat-Allah  and Masatoshi Imada  and Andreas M. Läuchli  and Guglielmo Mazzola  and Antonio Mezzacapo  and Andrew Millis  and Javier Robledo Moreno  and Titus Neupert  and Yusuke Nomura  and Jannes Nys  and Olivier Parcollet  and Rico Pohle  and Imelda Romero  and Michael Schmid  and J. Maxwell Silvester  and Sandro Sorella  and Luca F. Tocchio  and Lei Wang  and Steven R. White  and Alexander Wietek  and Qi Yang  and Yiqi Yang  and Shiwei Zhang  and Giuseppe Carleo },
title = {Variational benchmarks for quantum many-body problems},
journal = {Science},
volume = {386},
number = {6719},
pages = {296-301},
year = {2024},
doi = {10.1126/science.adg9774},
URL = {https://www.science.org/doi/abs/10.1126/science.adg9774},
eprint = {https://www.science.org/doi/pdf/10.1126/science.adg9774}}

@article{huse,
  title = {Simple Variational Wave Functions for Two-Dimensional Heisenberg Spin-\textonehalf{} Antiferromagnets},
  author = {Huse, David A. and Elser, Veit},
  journal = {Phys. Rev. Lett.},
  volume = {60},
  issue = {24},
  pages = {2531--2534},
  numpages = {0},
  year = {1988},
  month = {Jun},
  publisher = {American Physical Society},
  doi = {10.1103/PhysRevLett.60.2531},
  url = {https://link.aps.org/doi/10.1103/PhysRevLett.60.2531}
}

@article{rnn,
  title = {Leveraging recurrence in neural network wavefunctions for large-scale simulations of Heisenberg antiferromagnets on the triangular lattice},
  author = {Moss, M. Schuyler and Wiersema, Roeland and Hibat-Allah, Mohamed and Carrasquilla, Juan and Melko, Roger G.},
  journal = {Phys. Rev. B},
  volume = {112},
  issue = {13},
  pages = {134449},
  numpages = {21},
  year = {2025},
  month = {Oct},
  publisher = {American Physical Society},
  doi = {10.1103/nh89-6jmf},
  url = {https://link.aps.org/doi/10.1103/nh89-6jmf}
}

@article{sorella_sign,
  title = {Long-Range N\'eel Order in the Triangular Heisenberg Model},
  author = {Capriotti, Luca and Trumper, Adolfo E. and Sorella, Sandro},
  journal = {Phys. Rev. Lett.},
  volume = {82},
  issue = {19},
  pages = {3899--3902},
  numpages = {0},
  year = {1999},
  month = {May},
  publisher = {American Physical Society},
  doi = {10.1103/PhysRevLett.82.3899},
  url = {https://link.aps.org/doi/10.1103/PhysRevLett.82.3899}
}

@article{chen2024empowering,
  title={Empowering deep neural quantum states through efficient optimization},
  author={Chen, Ao and Heyl, Markus},
  journal={Nature Physics},
  volume={20},
  number={9},
  pages={1476--1481},
  year={2024},
  publisher={Nature Publishing Group UK London}
}

@article{spring,
title = {A Kaczmarz-inspired approach to accelerate the optimization of neural network wavefunctions},
journal = {Journal of Computational Physics},
volume = {516},
pages = {113351},
year = {2024},
issn = {0021-9991},
doi = {https://doi.org/10.1016/j.jcp.2024.113351},
url = {https://www.sciencedirect.com/science/article/pii/S0021999124005990},
author = {Gil Goldshlager and Nilin Abrahamsen and Lin Lin}
}

@article {carleo2017,
  title = {Solving the quantum many-body problem with artificial neural networks},
  author = {Carleo, G. and Troyer, M.},
  volume = {355},
  number = {6325},
  pages = {602--606},
  year = {2017},
  doi = {10.1126/science.aag2302},
  journal = {Science}
}

@article{Farnell2014,
  title = {Quantum $s=\frac{1}{2}$ antiferromagnets on Archimedean lattices: The route from semiclassical magnetic order to nonmagnetic quantum states},
  author = {Farnell, D. J. J. and G\"otze, O. and Richter, J. and Bishop, R. F. and Li, P. H. Y.},
  journal = {Phys. Rev. B},
  volume = {89},
  issue = {18},
  pages = {184407},
  numpages = {7},
  year = {2014},
  month = {May},
  publisher = {American Physical Society},
  doi = {10.1103/PhysRevB.89.184407},
  url = {https://link.aps.org/doi/10.1103/PhysRevB.89.184407}
}

@article{Yunoki2006,
   title = {Two spin liquid phases in the spatially anisotropic triangular Heisenberg model},
  author = {Yunoki, Seiji and Sorella, Sandro},
  journal = {Phys. Rev. B},
  volume = {74},
  issue = {1},
  pages = {014408},
  numpages = {31},
  year = {2006},
  month = {Jul},
  publisher = {American Physical Society},
  doi = {10.1103/PhysRevB.74.014408},
  url = {https://link.aps.org/doi/10.1103/PhysRevB.74.014408}
}

@article{Li2022,
  title = {Magnetization of the spin-$\frac{1}{2}$ Heisenberg antiferromagnet on the triangular lattice},
  author = {Li, Qian and Li, Hong and Zhao, Jize and Luo, Hong-Gang and Xie, Z. Y.},
  journal = {Phys. Rev. B},
  volume = {105},
  issue = {18},
  pages = {184418},
  numpages = {7},
  year = {2022},
  month = {May},
  publisher = {American Physical Society},
  doi = {10.1103/PhysRevB.105.184418},
  url = {https://link.aps.org/doi/10.1103/PhysRevB.105.184418}
}

@article{Hasik2024,
  author = {Hasik, J. and Corboz, P.},
  title = {Spiral iPEPS study of the spin-½ Heisenberg model on the triangular lattice},
  journal = {Phys. Rev. Lett.},
  volume = {133},
  pages = {176502},
  year = {2024}
}

@article{Angelucci1991,
  title = {Path-integral analysis of frustrated quantum Heisenberg models},
  author = {Angelucci, Antimo},
  journal = {Phys. Rev. B},
  volume = {44},
  issue = {13},
  pages = {6849--6857},
  numpages = {0},
  year = {1991},
  month = {Oct},
  publisher = {American Physical Society},
  doi = {10.1103/PhysRevB.44.6849},
  url = {https://link.aps.org/doi/10.1103/PhysRevB.44.6849}
}

@article{sandvik2026,
       author = {{Sandvik}, Anders W.},
        title = "{High-precision ground state parameters of the two-dimensional spin-1/2 Heisenberg model on the square lattice}",
      journal = {arXiv e-prints},
         year = 2026,
        month = jan,
archivePrefix = {arXiv},
       eprint = {2601.20189},
 primaryClass = {cond-mat.str-el},
       adsurl = {https://ui.adsabs.harvard.edu/abs/2026arXiv260120189S}
}

@article{robledomoreno2022,
   title={Fermionic wave functions from neural-network constrained hidden states},
   volume={119},
   ISSN={1091-6490},
   url={http://dx.doi.org/10.1073/pnas.2122059119},
   DOI={10.1073/pnas.2122059119},
   number={32},
   journal={Proceedings of the National Academy of Sciences},
   publisher={Proceedings of the National Academy of Sciences},
   author={Robledo Moreno, Javier and Carleo, Giuseppe and Georges, Antoine and Stokes, James},
   year={2022},
   month=aug }

@misc{roth2025,
      title={Superconductivity in the two-dimensional Hubbard model revealed by neural quantum states}, 
      author={Christopher Roth and Ao Chen and Anirvan Sengupta and Antoine Georges},
      year={2025},
      eprint={2511.07566},
      archivePrefix={arXiv},
      primaryClass={cond-mat.supr-con},
      url={https://arxiv.org/abs/2511.07566}, 
}

@article{xu2024,
   title={Coexistence of superconductivity with partially filled stripes in the Hubbard model},
   volume={384},
   ISSN={1095-9203},
   url={http://dx.doi.org/10.1126/science.adh7691},
   DOI={10.1126/science.adh7691},
   number={6696},
   journal={Science},
   publisher={American Association for the Advancement of Science (AAAS)},
   author={Xu, Hao and Chung, Chia-Min and Qin, Mingpu and Schollwöck, Ulrich and White, Steven R. and Zhang, Shiwei},
   year={2024},
   month=may }

@article{qin2020,
  title = {Absence of Superconductivity in the Pure Two-Dimensional Hubbard Model},
  author = {Qin, Mingpu and Chung, Chia-Min and Shi, Hao and Vitali, Ettore and Hubig, Claudius and Schollw\"ock, Ulrich and White, Steven R. and Zhang, Shiwei},
  collaboration = {Simons Collaboration on the Many-Electron Problem},
  journal = {Phys. Rev. X},
  volume = {10},
  issue = {3},
  pages = {031016},
  numpages = {18},
  year = {2020},
  month = {Jul},
  publisher = {American Physical Society},
  doi = {10.1103/PhysRevX.10.031016},
  url = {https://link.aps.org/doi/10.1103/PhysRevX.10.031016}
}

@article{savary2016,
   title={Quantum spin liquids: a review},
   volume={80},
   ISSN={1361-6633},
   url={http://dx.doi.org/10.1088/0034-4885/80/1/016502},
   DOI={10.1088/0034-4885/80/1/016502},
   number={1},
   journal={Reports on Progress in Physics},
   publisher={IOP Publishing},
   author={Savary, Lucile and Balents, Leon},
   year={2016},
   month=nov, pages={016502} }

@article{Fisher1989,
  title = {Universality, low-temperature properties, and finite-size scaling in quantum antiferromagnets},
  author = {Fisher, Daniel S.},
  journal = {Phys. Rev. B},
  volume = {39},
  issue = {16},
  pages = {11783--11792},
  numpages = {0},
  year = {1989},
  month = {Jun},
  publisher = {American Physical Society},
  doi = {10.1103/PhysRevB.39.11783},
  url = {https://link.aps.org/doi/10.1103/PhysRevB.39.11783}
}

@inproceedings{inductive_bias,
  title={The Need for Biases in Learning Generalizations},
  author={Tom Michael Mitchell},
  year={2007},
  url={https://api.semanticscholar.org/CorpusID:3237155}
}

@InProceedings{convit,
  title = 	 {ConViT: Improving Vision Transformers with Soft Convolutional Inductive Biases},
  author =       {D'Ascoli, St{\'e}phane and Touvron, Hugo and Leavitt, Matthew L and Morcos, Ari S and Biroli, Giulio and Sagun, Levent},
  booktitle = 	 {Proceedings of the 38th International Conference on Machine Learning},
  pages = 	 {2286--2296},
  year = 	 {2021},
  editor = 	 {Meila, Marina and Zhang, Tong},
  volume = 	 {139},
  series = 	 {Proceedings of Machine Learning Research},
  month = 	 {18--24 Jul},
  publisher =    {PMLR},
  url = 	 {https://proceedings.mlr.press/v139/d-ascoli21a.html},
}

@article{marshall1955,
  title = {Antiferromagnetism},
  author = {W. Marshall},
  journal = {Proceedings of the Royal Society of London. Series A, Mathematical and Physical Sciences},
  volume = {232},
  number = {1188},
  pages = {48--68},
  publisher = {The Royal Society},
  year = {1955},
  URL = {http://www.jstor.org/stable/99682},
}

@article{Ghioldi2015,
  title = {Magnons and excitation continuum in XXZ triangular antiferromagnetic model: Application to ${\text{Ba}}_{3}{\text{CoSb}}_{2}{\text{O}}_{9}$},
  author = {Ghioldi, E. A. and Mezio, A. and Manuel, L. O. and Singh, R. R. P. and Oitmaa, J. and Trumper, A. E.},
  journal = {Phys. Rev. B},
  volume = {91},
  issue = {13},
  pages = {134423},
  numpages = {8},
  year = {2015},
  month = {Apr},
  publisher = {American Physical Society},
  doi = {10.1103/PhysRevB.91.134423},
  url = {https://link.aps.org/doi/10.1103/PhysRevB.91.134423}
}

@article{Kaneko2014,
   title={Gapless Spin-Liquid Phase in an Extended Spin 1/2 Triangular Heisenberg Model},
   volume={83},
   ISSN={1347-4073},
   url={http://dx.doi.org/10.7566/JPSJ.83.093707},
   DOI={10.7566/jpsj.83.093707},
   number={9},
   journal={Journal of the Physical Society of Japan},
   publisher={Physical Society of Japan},
   author={Kaneko, Ryui and Morita, Satoshi and Imada, Masatoshi},
   year={2014},
   month=sep, pages={093707} 
}

@article{Heidarian2009,
  title = {Spin-$\frac{1}{2}$ Heisenberg model on the anisotropic triangular lattice: From magnetism to a one-dimensional spin liquid},
  author = {Heidarian, Dariush and Sorella, Sandro and Becca, Federico},
  journal = {Phys. Rev. B},
  volume = {80},
  issue = {1},
  pages = {012404},
  numpages = {4},
  year = {2009},
  month = {Jul},
  publisher = {American Physical Society},
  doi = {10.1103/PhysRevB.80.012404},
  url = {https://link.aps.org/doi/10.1103/PhysRevB.80.012404}
}

@article{Ghioldi2018,
    title = {Dynamical structure factor of the triangular antiferromagnet: Schwinger boson theory beyond mean field},
    author = {Ghioldi, E. A. and Gonzalez, M. G. and Zhang, Shang-Shun and Kamiya, Yoshitomo and Manuel, L. O. and Trumper, A. E. and Batista, C. D.},
    journal = {Phys. Rev. B},
    volume = {98},
    issue = {18},
    pages = {184403},
    numpages = {23},
    year = {2018},
    month = {Nov},
    publisher = {American Physical Society},
    doi = {10.1103/PhysRevB.98.184403},
    url = {https://link.aps.org/doi/10.1103/PhysRevB.98.184403}
}

@article{Goetze2016,
title={Ground-state properties of the triangular-lattice Heisenberg antiferromagnet with arbitrary spin quantum number s},
   volume={397},
   ISSN={0304-8853},
   url={http://dx.doi.org/10.1016/j.jmmm.2015.08.113},
   DOI={10.1016/j.jmmm.2015.08.113},
   journal={Journal of Magnetism and Magnetic Materials},
   publisher={Elsevier BV},
   author={Götze, O. and Richter, J. and Zinke, R. and Farnell, D.J.J.},
   year={2016},
   month=jan, pages={333–341}
}

@article{Capriotti1999,
  author    = {Capriotti, L. and Trumper, A. E. and Sorella, S.},
  title     = {Long-range N{\'e}el order in the triangular Heisenberg model},
  journal   = {Phys. Rev. Lett.},
  volume    = {82},
  pages     = {3899},
  year      = {1999},
  doi       = {10.1103/PhysRevLett.82.3899}
}

@article{White2007,
  title = {Ne\'el Order in Square and Triangular Lattice Heisenberg Models},
  author = {White, Steven R. and Chernyshev, A. L.},
  journal = {Phys. Rev. Lett.},
  volume = {99},
  issue = {12},
  pages = {127004},
  numpages = {4},
  year = {2007},
  month = {Sep},
  publisher = {American Physical Society},
  doi = {10.1103/PhysRevLett.99.127004},
  url = {https://link.aps.org/doi/10.1103/PhysRevLett.99.127004}
}

@article{Li2015,
  title = {Quasiclassical magnetic order and its loss in a spin-$\frac{1}{2}$ Heisenberg antiferromagnet on a triangular lattice with competing bonds},
  author = {Li, P. H. Y. and Bishop, R. F. and Campbell, C. E.},
  journal = {Phys. Rev. B},
  volume = {91},
  issue = {1},
  pages = {014426},
  numpages = {11},
  year = {2015},
  month = {Jan},
  publisher = {American Physical Society},
  doi = {10.1103/PhysRevB.91.014426},
  url = {https://link.aps.org/doi/10.1103/PhysRevB.91.014426}
}

@article{Zheng2006,
  title = {Excitation spectra of the spin-$\frac{1}{2}$ triangular-lattice Heisenberg antiferromagnet},
  author = {Zheng, Weihong and Fj\ae{}restad, John O. and Singh, Rajiv R. P. and McKenzie, Ross H. and Coldea, Radu},
  journal = {Phys. Rev. B},
  volume = {74},
  issue = {22},
  pages = {224420},
  numpages = {13},
  year = {2006},
  month = {Dec},
  publisher = {American Physical Society},
  doi = {10.1103/PhysRevB.74.224420},
  url = {https://link.aps.org/doi/10.1103/PhysRevB.74.224420}
}

@article{Richter2004,
    title = {Absence of magnetic order for the spin-half Heisenberg antiferromagnet on the star lattice},
      author = {Richter, J. and Schulenburg, J. and Honecker, A. and Schmalfu\ss{}, D.},
      journal = {Phys. Rev. B},
      volume = {70},
      issue = {17},
      pages = {174454},
      numpages = {6},
      year = {2004},
      month = {Nov},
      publisher = {American Physical Society},
      doi = {10.1103/PhysRevB.70.174454},
      url = {https://link.aps.org/doi/10.1103/PhysRevB.70.174454}
}

@article{Hawking1977,
  author = {Hawking, S. W.},
  title = {Zeta Function Regularization of Path Integrals in Curved Spacetime},
  journal = {Commun. Math. Phys.},
  volume = {55},
  pages = {133--148},
  year = {1977}
}

@book{Latticesums, 
doi = {10.1017/CBO9781139626804},
series={Encyclopedia of Mathematics and its Applications}, 
title={Lattice Sums Then and Now}, publisher={Cambridge University Press}, author={Borwein, J. M. and Glasser, M. L. and McPhedran, R. C. and Wan, J. G. and Zucker, I. J.}, 
year={2013}, 
collection={Encyclopedia of Mathematics and its Applications}
}

\end{document}


\title{Supplementary Information}
\maketitle
\tableofcontents
\section{Finite-Size Corrections in a Quantum Antiferromagnet}

\subsection{Notation}
Given a two-dimensional lattice of $L \times L$ spins on a torus generated by primitive lattice vectors ${\bm a}_1$ and ${\bm a_2}$, we view the torus as the unit cell of an infinite lattice with lattice vectors ${\bm A}_1=L{\bm a}_1$ and ${\bm A}_2=L{\bm a}_2$ and lattice points $\bm{R} = n_1 \bm{A}_1 + n_2 \bm{A}_2$ with $n_1, n_2 \in \mathbb{Z}$.
The supercell area is
\begin{equation}\label{eq:volume}
V = |\bm{A}_1 \times \bm{A}_2| = L^2 |\bm{a}_1 \times \bm{a}_2|
\end{equation}

The corresponding reciprocal lattice vectors are defined as
\begin{equation}
\bm{B}_1 = \frac{2\pi}{V}\,(\bm{A}_2 \times \hat{\bm{z}})
\qquad
\bm{B}_2 = \frac{2\pi}{V}\,(\hat{\bm{z}} \times \bm{A}_1) \ ,
\end{equation}

and satisfy $\bm{A}_i \cdot \bm{B}_j = 2\pi\,\delta_{ij}$. Allowed reciprocal vectors are therefore $\bm{G} = m_1 \bm{B}_1 + m_2 \bm{B}_2$ with $m_1, m_2 \in \mathbb{Z}$.
It is convenient to introduce primitive reciprocal vectors $\bm{b}_i$ via $\bm{B}_i = \tfrac{2\pi}{L}\,\bm{b}_i$ such that $\bm{a}_i \cdot \bm{b}_j = \delta_{ij}$.
Then the norm of $\bm{G}$ can be written as
\begin{equation}\label{eq:module_G}
|\bm{G}| = \frac{2\pi}{L}\,|\bm{m}|_g \qquad |\bm{m}|_g = \sqrt{\bm{m}^T g\,\bm{m}} \ ,
\end{equation}
where the metric tensor  in reciprocal space is defined by $g_{ij} = \bm{b}_i \cdot \bm{b}_j$. The reciprocal-cell area satisfies
\begin{equation}\label{eq:det_g}
\sqrt{\det g} = |\mathbf{b}_1\times \mathbf{b}_2| = \frac{1}{|\mathbf{a}_1\times \mathbf{a}_2|} \ ,
\end{equation}
so that \cref{eq:volume} can equivalently be written as $V = L^2/\sqrt{\det g}$.

\subsection{Ground-State Energy}
In this section we extract the leading finite-size correction to the ground-state energy of a two-dimensional quantum antiferromagnet on a torus.
In the continuum limit, the ground-state energy density at $T=0$ for the triangular antiferromagnet is
\begin{equation}
e_0 = \left( \sum_{\ell=1}^{3} \frac{c_\ell }{2} \right) \int_{|{\mathbf{k}}| < \Lambda} \frac{d^2 k}{4 \pi^2}  |{\mathbf{k}}| \ ,
\end{equation}
which depends on the ultraviolet cutoff $\Lambda$.
Following Ref.~\cite{Fisher1989}, we compute the finite-size corrections, which are independent of the cutoff. By dimensional analysis and scaling, the energy on a torus of volume $V\propto L^2$ admits the scaling form
\begin{equation}\label{eq:en_scaling_form}
\frac{E}{L^2} = e_0 + \left( \sum_{\ell=1}^{3} \frac{c_\ell }{2} \right) \frac{\mathcal{C}_1}{L^{3}} + \dots \ .
\end{equation}
Here $e_0$ depends on $\Lambda$ and is non-universal, whereas $\mathcal{C}_1$ is universal: it depends only on the modular parameter of the torus and is {\it independent\/} of the underlying lattice on which the spins sit.

In the discrete setting, the energy per unit volume can be expressed (after applying the Poisson summation formula) as
\begin{equation}
\frac{E}{V} = \left( \sum_{\ell=1}^{3} \frac{c_\ell }{2} \right) \frac{1}{V} \sum_{{\bm G}} |{\bm G}| 
 = \left( \sum_{\ell=1}^{3} \frac{c_\ell }{2} \right) \int_{\mathbb{R}^2} \frac{d^2 k}{4 \pi^2} \sum_{{\bm R}} e^{i {\mathbf{k}}\cdot{\bm R}} |{\mathbf{k}}| \ .
\end{equation}
Performing the Fourier transform of $|{\mathbf{k}}|$ and exploiting \cref{eq:det_g}, one obtains:
\begin{equation}\label{eq:en_scaling2}
    \frac{E}{L^2}= e_0 - \frac{1}{\sqrt{\det g}}\left( \sum_{\ell=1}^{3} \frac{c_\ell }{2} \right) \sum_{{\bm R}\neq {\bm 0}} \frac{1}{2 \pi |{\bm R}|^3} \ .
\end{equation}
For the triangular lattice we choose $\bm a_1 = (1,0)$ and $\bm a_2 = (-\tfrac12,\tfrac{\sqrt3}{2})$. Then $|\bm R|^3 = L^3\,(n_1^2 + n_2^2-n_1n_2)^{3/2}$ and we find
\begin{equation}
    \frac{E}{L^2}= e_0 - \left( \sum_{\ell=1}^{3} \frac{c_\ell }{2} \right) \left[\frac{S_{\triangle} }{2\pi\sqrt{\det g}} \right] \frac{1}{L^3} \ ,
\end{equation}
where we have defined $S_{\triangle}=\sum_{{(n_1,n_2)} \neq {\bm 0}} (n_1^2 + n_2^2 -n_1n_2)^{-3/2}$. The last expression is of the form of \cref{eq:en_scaling_form} with $\mathcal{C}_1=-S_{\triangle}/(2\pi\sqrt{\det g})$ where $S_{\triangle}=6\,
\zeta\!\left(\tfrac{3}{2}\right)\,
L\!\left(\chi_{-3},\tfrac{3}{2}\right)=11.03417573$. Here, $\zeta(\cdot)$ is the Riemann zeta function and $L(\chi_{-3}, \cdot)$ is the Dirichlet $L$-function associated with the Dirichlet character $\chi_{-3}$ \cite{Latticesums}. 
Finally, we obtain
\begin{equation}
    \boxed{\mathcal{C}_1 = -\frac{S_{\triangle}}{2\pi\sqrt{\det g}} = -1.52086497}\ .
\end{equation}

The analogous calculation can be performed for the square lattice with $\bm a_1 = (1,0)$ and $\bm a_2 = (0,1)$, velocities $c_1=c_2=c$, and $\sqrt{\det g}=1$. In this case \cite{Latticesums}
\begin{equation}
S_{\square}=\sum_{(n_1,n_2)\neq \bm{0}} \frac{1}{(n_1^2 + n_2^2)^{3/2}}=4\,\zeta\!\left(\tfrac{3}{2}\right)\,\beta\!\left(\tfrac{3}{2}\right)= 9.03362168 \ ,
\end{equation}
where $\beta(\cdot)$ is the Dirichlet beta function. We obtain $\mathcal{C}_1= -S_{\square}/(2\pi)=-1.437745$, in agreement with Refs.~\cite{Fisher1989, sandvik1997, sandvik2026}.

\subsection{Ground-State Magnetization}
In this section we determine the leading finite-size correction to the ground-state magnetization of a two-dimensional quantum antiferromagnet on a torus. As stated in the main text, the magnetization on the triangular lattice is characterized by two orthogonal unit vectors $\bm{n}_1$ and $\bm{n}_2$, which are related to $z_{\alpha}$ by
\begin{equation}
(n_1)^a + i\,(n_2)^a = z_{\gamma}\,\varepsilon_{\gamma\alpha}\,\sigma^a_{\alpha\beta}\,z_{\beta}, \label{C1}
\end{equation}
We solve the constraint $\sum_{\alpha} |z_{\alpha}|^2 = 1$ by parametrizing
\begin{align}
z_1 = \sqrt{1 - \pi_1^2 - \pi_2^2 - \pi_3^2} + i\pi_1, \qquad z_2 = \pi_2 + i\pi_3, \label{C2}
\end{align}
where $\pi_{\ell}$ are three real fields ($\ell=1,2,3$).
To second order in $\pi_{\ell}$, the relation in \cref{C1} implies
\begin{equation}
\bm{n}_1=
\begin{pmatrix}
1-2\left(\pi_1^2+\pi_2^2\right)\\
-2\pi_1-2\pi_2\pi_3\\
-2\pi_2+2\pi_1\pi_3
\end{pmatrix}
+\cdots,
\qquad
\bm{n}_2=
\begin{pmatrix}
2\pi_1-2\pi_2\pi_3\\
1-2\left(\pi_1^2+\pi_3^2\right)\\
-2\pi_3-2\pi_1\pi_2
\end{pmatrix}
+\cdots.
\end{equation}
From this parametrization, and the action for $\pi_\ell$ in the main text, we find that
\begin{equation}
    c_1 = c_{\parallel},
    \qquad
    c_2 = c_3 = c_{\perp},
\end{equation}
as $\pi_1$ is the in-plane mode, while $\pi_{2,3}$ are the two out-of-plane modes.

We evaluate the structure factor on a torus of volume $V$,
\begin{align}
\frac{\mathcal{M}^2}{\mathcal{M}_0^2}
&= \frac{1}{V}\int d^2x\,\langle \bm{n}_1(x,\tau)\cdot \bm{n}_1(0,\tau)\rangle \\
&= 1 + \frac{4}{V}\sum_{\ell=1}^{2}\int\frac{d\omega}{2\pi}\left(
\frac{1}{m_{\ell}\omega^2}-\sum_{\bm{G}}\frac{1}{m_{\ell}(\omega^2+c_{\ell}^2|\bm{G}|^2)}
\right)\\
&= 1-\frac{2}{V}\left(\sum_{\ell=1}^{2}\frac{1}{m_{\ell}c_{\ell}}\right)\sum_{\bm{G}\neq 0}\frac{1}{|\bm{G}|}.
\end{align}
Subtracting the infinite-volume answer, we have
\begin{equation}\label{eq:magnetization}
\mathcal{M}^2-\mathcal{M}_{0}^2
= - \mathcal{M}_0^2 \left(\sum_{\ell=1}^{2}\frac{2}{m_{\ell}c_{\ell}}\right)
\left(
\frac{1}{V}\sum_{\bm{G}\neq 0}\frac{1}{|\bm{G}|}
- \int_{|\bm{k}| \le \Lambda}\frac{d^2k}{4\pi^2}\,\frac{1}{|\bm{k}|}
\right).
\end{equation}
The limit $\Lambda\to +\infty$ should exist after the subtraction; in practice, one may introduce a smooth cutoff. This yields a result of the form
\begin{equation}
\mathcal{M}^2(L)=\mathcal{M}_{0}^2
+ \mathcal{M}_0^2 \left(\sum_{\ell=1}^{2}\frac{2}{m_{\ell}c_{\ell}}\right)\frac{\mathcal{C}_2}{L} \ ,
\end{equation}
similar to \cref{eq:en_scaling_form}. Here $\mathcal{C}_2$ is a universal number which depends only on the modular parameter of the torus.

To extract $\mathcal{C}_2$ we evaluate the difference between the discrete reciprocal-lattice sum and the corresponding continuum integral in \cref{eq:magnetization}. Each term diverges, but their difference is finite.

\begin{figure}[t]
    \begin{center}
\centerline{\includegraphics[width=0.6\columnwidth]{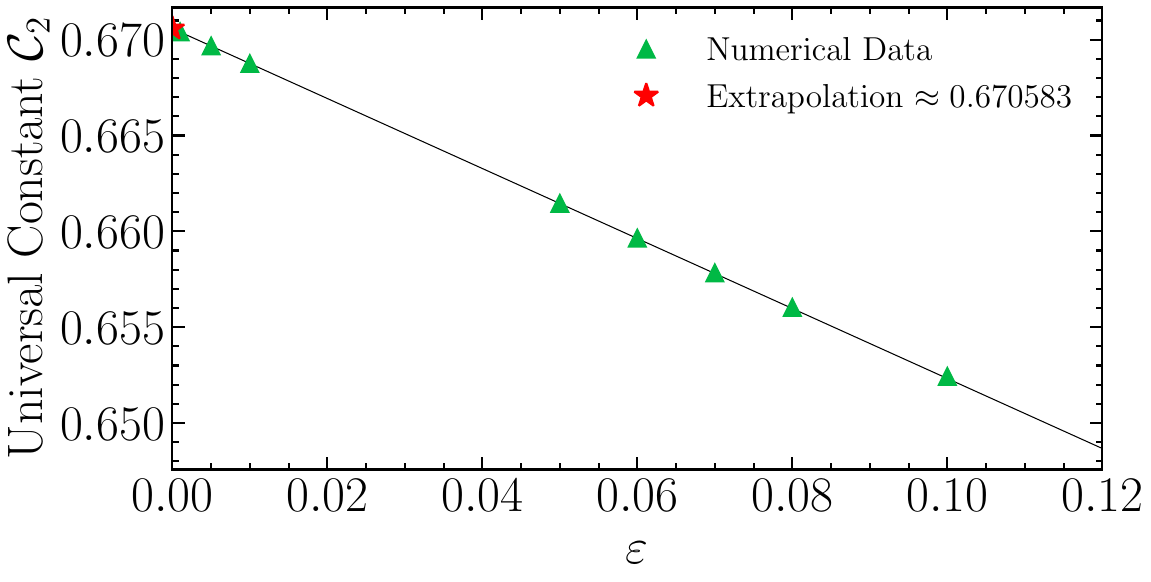}}
        \caption{Finite-size contributions $\mathcal{C}_2$ of the ground-state magnetization extracted from the regulated difference between the discrete reciprocal-lattice sum and the corresponding continuum integral, shown as a function of the regulator $\varepsilon$.\label{fig:finite_contributions}}
    \end{center}
\end{figure}

We use two complementary regularizations to validate the result. First, we introduce an exponential regulator~\footnote{This is equivalent to a hard momentum cutoff $\Lambda$ in \cref{eq:magnetization}, but it improves numerical convergence.} and define
\begin{equation}
\Delta_{\eta}(L)
\equiv \frac{1}{V}\sum_{\mathbf{G}\neq \mathbf{0}} \frac{e^{-\eta\,|\mathbf{G}|}}{|\mathbf{G}|}
- \int_{\mathbb{R}^2}\frac{d^2\mathbf{k}}{4\pi^2}\,\frac{e^{-\eta\,|\mathbf{k}|}}{|\mathbf{k}|} \ .
\end{equation}
The continuum integral can be evaluated explicitly using polar coordinates in $\mathbf{k}$.
\begin{equation}
\int_{\mathbb{R}^2}\frac{d^2\mathbf{k}}{4\pi^2}\,\frac{e^{-\eta\,|\mathbf{k}|}}{|\mathbf{k}|}
= \frac{1}{2\pi\eta} \ .
\end{equation}
Introducing the dimensionless regulator $\varepsilon \equiv (2\pi\eta)/L$ and exploiting \cref{eq:det_g} and \cref{eq:module_G}, we arrive at the following dimensionless expression:
\begin{equation}\label{eq:delta_eps}
\Delta_{\varepsilon}(L)
\equiv \frac{1}{2\pi L}\left[\sqrt{\det g}
\sum_{\mathbf{m}\neq \mathbf{0}}\frac{e^{-\varepsilon|\mathbf{m}|_{g}}}{|\mathbf{m}|_{g}}
- \frac{2\pi}{\varepsilon}
\right] \ .
\end{equation}
For the small-$\varepsilon$ expansion we expect
\begin{equation}
\sum_{\mathbf{m}\neq \mathbf{0}}\frac{e^{-\varepsilon|\mathbf{m}|_{g}}}{|\mathbf{m}|_{g}
}
\underset{\varepsilon \rightarrow 0}{\sim}
\frac{2\pi}{\varepsilon\,\sqrt{\det g}} + C + O(\varepsilon) \ .
\end{equation}
Consequently, we can extract the finite part as $\varepsilon\to 0$:
\begin{equation}\label{eq:finite_cont_LR1}
\Delta_{\varepsilon}(L)
\underset{\varepsilon \rightarrow 0}{\sim}
\frac{1}{2\pi L}\left[
\sqrt{\det g}\left(\frac{2\pi}{\varepsilon\,\sqrt{\det g}} + C + O(\varepsilon)\right)
- \frac{2\pi}{\varepsilon}
\right]
= \frac{1}{2\pi L}\,\sqrt{\det g}\,C + O(\varepsilon) \ .
\end{equation}
In the end, the finite contribution can be rewritten as
\begin{equation}\label{eq:finite_cont_LR2}
\mathcal{C}_2 \equiv -\frac{\sqrt{\det g}\ C}{2\pi} \ .
\end{equation}
The latter can be computed numerically by evaluating the difference between the discrete sum and the integral for several values of $\varepsilon$ and then extrapolating to $\varepsilon \to 0$, yielding $\mathcal{C}_2 \approx 0.670583$ (see \cref{fig:finite_contributions}).

As an alternative approach, we use \textit{zeta-function regularization} \cite{Hawking1977}, which yields an analytic result. We start by defining the lattice sum appearing in the \cref{eq:delta_eps}
\begin{equation}\label{eq:series_magn}
    Z_g = \sqrt{\det g} \sum_{\bm{m} \neq \bm{0}} \frac{1}{|\bm{m}|_g} = \sum_{(m_1,m_2) \neq \bm 0} \frac{1}{(m_1^2 + m_2^2 + m_1m_2)^{1/2}} \ .
\end{equation}
Here we used $|\bm{m}|_g = \tfrac{2}{\sqrt{3}}(m^2 + n^2 + mn)^{1/2}$ and $\sqrt{\det g}=\tfrac{2}{\sqrt{3}}$.

The series in \cref{eq:series_magn} is divergent. However, it can be defined by analytic continuation via the \textit{triangular-lattice Epstein zeta function} \cite{Latticesums},
\begin{equation}
Z_{\triangle}(s)=\sum_{(m_1,m_2) \neq \bm 0}\frac{1}{\left(m_1^2+m_1m_2+m_2^2\right)^s} \ ,
\qquad
s \in \mathbb{C}, \ \operatorname{Re}(s)>1\,.
\end{equation}
This function admits an analytic continuation and a functional equation.
One finds the factorization
\begin{equation}\label{eq:z_triangular}
Z_{\Delta}(s)=6\,\zeta(s)\,L(\chi_{-3},s) \ ,
\end{equation}
where $\zeta(\cdot)$ is the Riemann zeta function and $L(\chi_{-3},\cdot)$ is the Dirichlet $L$-function associated with the Dirichlet character $\chi_{-3}$ \cite{Latticesums}.
Notice that \cref{eq:z_triangular} admits a meromorphic continuation to the complex plane with a single simple pole at $s=1$. One finds by analytic continuation that, for $s=1/2$
\begin{equation}\label{eq:result_triang}
Z_{\Delta}\!\left(\tfrac{1}{2}\right)=6\,\zeta\!\left(\tfrac{1}{2}\right)\,L\!\left(\chi_{-3}, \tfrac{1}{2}\right) = -4.21342263613690689003 \ .
\end{equation}
With this procedure we can directly extract the finite part of the series without subtracting divergent contribution \cite{Hawking1977} and obtaining
\begin{equation}
    \boxed{\mathcal{C}_2 = -\frac{Z_{\triangle}(\tfrac12)}{2\pi} = 0.670587039} \ .
\end{equation}
Notice that the result obtained with the latter approach matches the one obtained with the exponential cutoff regularization (see \cref{fig:finite_contributions}).

The analogous calculation can be performed for the square lattice defining 
\begin{equation}
    Z_{\square}(s)=\sum_{{(m_1,m_2)} \neq {\bm 0}} \frac{1}{(m_1^2 + m_2^2)^{s}} \ ,
\end{equation}
which can be derived analytically as $Z_{\square}(s)=4\,\zeta\!\left(s\right)\,
\beta\!\left(s\right)$, and evaluated at $s=\tfrac{1}{2}$ obtaining ${Z_{\square}(\tfrac{1}{2})=-3.900264920001}$ and consequently ${\mathcal{C}_2=-Z_{\square}(s)}/(2\pi)=0.62074644$ in agreement with Refs.~\cite{sandvik1997, sandvik2026}.

\section{Numerical Results from Linear Spin-Wave Theory}

We consider the spin-$\tfrac12$ Heisenberg antiferromagnet on the triangular lattice. In this section we report the results obtained with linear spin wave theory following Ref.~\cite{chubukov1999} and setting throughout $S=\tfrac12,J=1,a=1$. The two independent spin-wave velocities are
\begin{equation}
c^0_\parallel
= \frac{3\sqrt{3}}{2} J S a
\qquad
c^0_\perp
= \frac{3\sqrt{3}}{2\sqrt{2}} J S a \ .
\end{equation}

Including the $1/S$ corrections, the renormalized velocities read
\begin{equation}
c_{\text{SW}\parallel}
= c^0_\parallel \left(1-\frac{0.115}{2S}\right)
\qquad
c_{\text{SW}\perp}
= c^0_\perp \left(1+\frac{0.083}{2S}\right) \ .
\end{equation}

For $S=\tfrac12$ this gives
\begin{equation}
     c_{\text{SW}\parallel} = 1.1496487235238424 \qquad c_{\text{SW}\perp} = 0.9947990217878181 \ .
\end{equation}

The uniform susceptibilities are given by
\begin{equation}
\chi_{\text{SW}\perp} = \frac{2}{9\sqrt{3} J a^2} \left(1-\frac{0.285}{2S}\right)
\qquad
\chi_{\text{SW}\parallel} = \frac{2}{9\sqrt{3} J a^2} \left( 1-\frac{0.449}{2S}\right) \ .
\end{equation}
For $S=\tfrac12$ one finds
\begin{equation}
    \chi_{\text{SW}\parallel} = 0.07069333296077418
\qquad \chi_{\text{SW}\perp} = 0.09173454277124056 \ .
\end{equation}

The corresponding masses entering the nonlinear $\sigma$-model are $m_1 = 4\chi_\parallel$, $m_2 = m_3 = 4\chi_\perp$ and using the relations $c_1 = c_\parallel$, $c_2=c_3=c_\perp$ the coefficient entering the finite-size correction to the ground-state energy is
\begin{equation}
\bar{c}
= c_\parallel + 2c_\perp
= 3.1392467670994786 \ .
\end{equation}

Similarly, the coefficient controlling the finite-size scaling of the magnetization is given by
\begin{equation}
    \frac{1}{\overline{c\chi}_{\text{SW}}}= \frac{1}{2}\left(\frac{1}{c_{\text{SW}\parallel}\chi_{\text{SW}\parallel}} + \frac{1}{c_{\text{SW}\perp}\chi_{\text{SW}\perp}}\right) = 11.631148207320964 \ ,
\end{equation}
and as a result ${\overline{c\chi}_{\text{SW}}} = 0.08597603453$.

\bibliography{ref}